\documentclass[%
superscriptaddress,
nofootinbib,
nobibnotes,
amsmath,amssymb,
aps,
floatfix,
]{revtex4-2}

\usepackage{placeins}
\usepackage{booktabs}
\usepackage{dcolumn}
\usepackage{array}
\usepackage{longtable}
\usepackage{float}
\usepackage{hyphenat}
\usepackage{natbib}
\usepackage{hyperref}

\usepackage{amsmath,amsfonts,amsthm,amssymb}
\usepackage{bm,graphicx,graphics,color}
\usepackage{epsf,epsfig}
\usepackage[all]{xy}
\newcommand*{\B}[1]{\ifmmode\bm{#1}\else\textbf{#1}\fi}

\def\bx{\mathbf{x}}

\def\bv{\mathbf{v}}
\def\bV{\mathbf{V}}

\def\bp{\mathbf{p}}

\def\no{\nonumber}
\def\lb{\label}
\def\be{\begin{equation}}
\def\ee#1{\label{#1}\end{equation}}
\newcommand{\ben}{\begin{eqnarray}}
\newcommand{\een}{\end{eqnarray}}

\begin{document}

\title{Boltzmann equation in the $2{\frac12}$-post-Newtonian approximation\footnote{Dedicated to Ingo M\"uller on the occasion of his ninetieth  birthday}}

\author{Gilberto M. Kremer}
\email{kremer@fisica.ufpr.br}
\affiliation{Departamento de F\'{i}sica, Universidade Federal do Paran\'{a}, Curitiba 81531-980, Brazil}

%%%%%%%%%%%%%%%%%%%%%%%%%%%%%%%%%%%%%%%%%%%%%%%%%%%%%%%%%%%%%%%
\begin{abstract}
Within the framework of the post-Newtonian $2\frac12$ approximation theory, a kinetic theory for relativistic gases in the presence of gravitational fields is developed. The  Boltzmann equation and  the equilibrium Maxwell--J\"uttner distribution function are determined up to $1/c^7$--order, which are used to calculate  the components of the particle four-flow and energy-momentum tensor and to find the  Eulerian hydrodynamic equations for the mass, mass-energy, and momentum densities in the $2\frac12$--post-Newtonian approximation.   The energy conservation law follows from the hydrodynamic equation for the total energy density, which is a combination of the hydrodynamic equations for the mass and the mass-energy densities. Here, the  total energy conservation law is derived in the $1\frac12$--post-Newtonan approximation, since we have to know the components of the metric tensor $g_{ij}$ in the $3\frac12$--th post-Newtonian approximation to obtain the energy conservation law in the $2\frac12$--th post-Newtonian approximation.
\end{abstract}
\maketitle 
%%%%%%%%%%%%%%%%%%%%%%%%%%%%%%%%%%%%%%%%%%%%%%%%%%%%

\section{Introduction}
\lb{s.1}

The determination of  post-Newtonian (PN)  approximations  from  Einstein's field equations goes back to the seminal paper proposed by Einstein, Infeld and Hoffmann  in 1938 \cite{Eins}. The method of successive approximations in even powers of the inverse of the speed of light $c$ was developed in a series of papers by Chandrasekhar et. al \cite{Ch0,Ch1,Ch2,Ch3,Nu} (see also the books \cite{Fock,Wein,Will,GK1} for the description of the method). 

For the expansion in odd powers of the inverse of the speed of light, the Sommerfeld radiation condition at infinity and the harmonic de Donder coordinate conditions were used \cite{Ch4,Ch5,And} to solve the equivalent form of Einstein's field equations due to Landau and Lifschitz \cite{LL} and which makes use of the conservation of the energy-momentum complex.  %In the works \cite{Ch4,Ch5} the methodology was the one proposed by Chandrasekhar, which makes use of some coordinate gage conditions, while in the work \cite{And} the solutions of Einstein's field equations are obtained from a slow motion expansion in inverse powers of the speed of light.

In the works of Chandrasekhar et. al \cite{Ch0,Ch1,Ch2,Ch3,Nu,Ch4,Ch5} one can find a complete phenomenological description  of the hydrodynamic equations, conservation equations and  equations for the gravitational potentials up to the $2\frac12$--PN approximation. An important  outcome   from the $2\frac12$--PN approximation is that the  mean value of the energy rate "represents a secular decrease of the integrated energy of the system, a result that is in exact agreement with the rate of emission of gravitational radiant energy predicted by the linearized theory of gravitational radiation \cite{Ch4,Ch5}." In these works it was shown that the average over a long time of the total energy density in the  $2\frac12$--PN approximation
$\mathfrak{E}_{(2.5)}$ is not conserved, but given by
\ben\no
\Bigg\langle \frac{d}{dt}\int \mathfrak{E}_{(2.5)}d\bx\Bigg\rangle=-\frac{G}{45c^5}\Bigg\langle \Bigg\vert\frac{d^3}{dt^3}\left(3 I_{ij}-I_{kk}\delta_{ij}\right)\Bigg\vert^2\Bigg\rangle,
\een 
where $I_{ij}=\int_V\rho x_ix_jd\bx$ denotes the moment of inertia tensor. The   moment of inertia tensor is related to the gravitational potentials in the $1\frac12$--PN and $2\frac12$--PN approximations (see equations (\ref{q1}) and (\ref{q2}) below).

The determination of the collisionless Boltzmann equation within the framework of the first  and second  PN approximations was the subject of the works  \cite{Rez, Ped} and \cite{GK1,GK2}, respectively. The extension of the PN Boltzmann equation to include the non-equilibrium contributions was also analyzed in \cite{GK3,GK4}. Some applications of the PN Boltzmann equation can be found in the works \cite{GK5,GK6,GK7,GK8} and the references therein.   

The aim of the present work is to extend the tools and techniques of the $2\frac12$--PN approximation to a kinetic theory of gases described by the Boltzmann equation, which is the space-time evolution equation for the one-particle distribution function,  derived from the evolution equation for the one-particle distribution function with respect to the proper time along the world line of a gas particle. As was shown in the literature, the post-Newtonian kinetic equation is an important tool to analyze self-gravitating problems and the influence of the new gravitational potentials from the $2\frac12$--post--Newtonian  theory is  a topic which deserves some attention.  

The equilibrium state of a relativistic gas is described by the one-particle  Maxwell--J\"uttner distribution function, which is obtained from the collision term of the relativistic Boltzmann equation by assuming that the number of particles that enter and leave the volume element in the phase space of the particles are the same (see, e.g. \cite{CK}). The first and second PN approximations to the Maxwell--J\"uttner distribution function were derived in\cite{KRW,GK1,GK2}. Here  we follow the same methodology of these works to obtain the $2\frac12$--PN approximation of the Maxwell--J\"uttner distribution function up to the $1/c^7$--order.  

Following the usual scheme of the kinetic theory of gases,  from knowledge of the Maxwell--J\"uttner distribution function the components of the particle four-flow and of the energy-momentum tensor can be derived from their definitions given in terms of the one-particle distribution function. The expressions  obtained for the components correspond to those that follow from a phenomenological theory based on their decompositions in terms of the hydrodynamic four-velocity.  

The hydrodynamic equations for the mass, mass-energy  and momentum densities for an Eulerian fluid in the $2\frac12$--PN approximation are derived from the Boltzmann equation, which correspond to the phenomenological conservation equations  of the particle four-flow and of the energy-momentum tensor.    % The Eulerian hydrodynamic equations derived from the post-Newtonian Boltzmann equation correspond to those obtained from the phenomenological theory which follow from the \cite{ChNu}. By neglecting the relativistic corrections the hydrodynamic equations for the mass  and mass-energy densities coincide and  correspond to the Newtonian continuity equation. However, their difference  leads to the hydrodynamic equation for the internal energy density, which is an expression in the first post-Newtonian approximation.  This result is compatible with  energy conservation law in the post-Newtonian theory \cite{Ch1,Ch2,ChNu}, since the first post-Newtonian expression for the energy conservation law follows only from the knowledge of the second post-Newtonian approximation. 

From the combination of the hydrodynamic equations for mass and mass-energy, the corresponding hydrodynamic equation for the total energy density follows, from which the energy conservation law can be derived. As  Chandrasekhar \cite{Ch0,Ch1,Ch2,Ch3,Ch4,Ch5} pointed, an important feature of the PN approximations is that the energy conservation law is determined in its  $n$--th PN approximation, once  the components of the metric tensor in the $(n+1)$--th PN approximation are known. Here  only  the energy conservation law in the $1\frac12$--PN approximation was derived and it is shown that it is the same as the one of the first PN approximation, i.e. there is no contribution of the gravitational potential of the $1\frac12$--PN approximation to the energy conservation law. For the determination of the energy conservation law in the $2\frac12$--PN approximation it is necessary to know the components of the metric tensor in the $3\frac12$--PN approximation (see \cite{Ch4,Ch5}).

The paper  is outlined as follows: in Section \ref{s.2} we introduce the main outcomes from the $2\frac12$--PN approximation theory. The  $2\frac12$--PN Boltzmann equation is derived in Section \ref{s.3}, the  Maxwell--J\"uttner distribution function is determined in Section \ref{s.4}, while the derivation of the energy-momentum tensor components and the Eulerian hydrodynamic equations   are the subject of  Sections  \ref{s.5} and \ref{s.6}, respectively. The energy conservation law is analyzed in Section \ref{s.7} and the conclusions of the work are given in the last section. 
The notations used are: Greek indices take the values 0,1,2,3 and Latin indices the values 1,2,3. The semicolon denotes the covariant differentiation, the indices of Cartesian tensors will be written as subscripts, the summation convention over repeated  indices will be assumed, and the partial differentiation will be denoted by $\partial/\partial x^i$.

\section{Preliminaries}
\lb{s.2}

The starting point of the post-Newtonian approximation in odd powers of the inverse of the light speed $c$ is the use of the  Sommerfeld radiation condition at infinity and the harmonic de Donder coordinate conditions \cite{Ch4,And}, namely  ${\mathfrak{g}^{\mu\nu}}_{,\nu}=0$, where $\mathfrak{g}^{\mu\nu}=\sqrt{-g}g^{\mu\nu}$ is the metric tensor density, $g^{\mu\nu}$ are the components of the gravitational field and  $g$ its determinant. 

The harmonic de Donder coordinate conditions are used to solve the equivalent form of Einstein's field equations due to Landau and Lifschitz \cite{LL} in terms of the energy-momentum complex $\Theta^{\mu\nu}$, which reads
\ben\lb{ON1}
\Theta^{\mu\nu}=\frac{c^4}{16\pi G}\left[\mathfrak{g}^{\mu\nu}\mathfrak{g}^{\rho\sigma}-\mathfrak{g}^{\mu\rho}\mathfrak{g}^{\nu\sigma}\right]_{,\rho\sigma},
\een
where $G$ is the gravitational constant.  The energy-momentum complex  is a function of the energy-momentum tensor $T^{\mu\nu}$ (source of the gravitational field)   and of the pseudo-tensor $t^{\mu\nu}$ and given by
$\Theta^{\mu\nu}=(-g)\left(T^{\mu\nu}+t^{\mu\nu}\right)$, while 
the pseudo-tensor is written in terms of the Christoffel symbols ${\Gamma^\tau}_{\epsilon\lambda}$ by (see \cite{LL})
\ben\no
t^{\mu\nu}=\frac{c^4}{16\pi G}\Big(g^{\mu\nu}g^{\sigma\lambda}-g^{\mu\lambda}g^{\sigma\nu}\Big)\Big[{\Gamma^\tau}_{\tau\sigma}{\Gamma^\epsilon}_{\epsilon\lambda}+{\Gamma^\tau}_{\epsilon\lambda}{\Gamma^\epsilon}_{\tau\sigma}-2{\Gamma^\tau}_{\tau\epsilon}{\Gamma^\epsilon}_{\sigma\lambda}\Big]+g^{\mu\lambda}g^{\sigma\epsilon}\Big[{\Gamma^\tau}_{\tau\lambda}{\Gamma^\nu}_{\sigma\epsilon}+{\Gamma^\tau}_{\sigma\epsilon}{\Gamma^\nu}_{\tau\lambda}-{\Gamma^\tau}_{\tau\epsilon}{\Gamma^\nu}_{\sigma\lambda}
\\\lb{ON3}
-{\Gamma^\tau}_{\sigma\lambda}{\Gamma^\nu}_{\tau\epsilon}\Big]+g^{\nu\lambda}g^{\sigma\epsilon}\Big[{\Gamma^\tau}_{\tau\lambda}{\Gamma^\mu}_{\sigma\epsilon}+{\Gamma^\tau}_{\sigma\epsilon}{\Gamma^\mu}_{\tau\lambda}-{\Gamma^\tau}_{\tau\epsilon}{\Gamma^\mu}_{\sigma\lambda}
-{\Gamma^\tau}_{\sigma\lambda}{\Gamma^\mu}_{\tau\epsilon}\Big]+g^{\lambda\epsilon}g^{\sigma\tau}\Big[{\Gamma^\mu}_{\tau\lambda}{\Gamma^\nu}_{\sigma\epsilon}-{\Gamma^\mu}_{\lambda\epsilon}{\Gamma^\nu}_{\tau\sigma}\Big].\qquad
\een

The vanishing divergence of the energy-momentum complex ${\Theta^{\mu\nu}}_{,\nu}=0$, implies the vanishing  covariant divergence 
of the energy-momentum tensor ${T^{\mu\nu}}_{;\nu}=0$. 

The energy-momentum tensor -- the source of the gravitational field -- is that of a relativistic perfect fluid, written in terms of its pressure $p$ and energy density $\epsilon$:
\ben
T^{\mu\nu}=(\epsilon+p)\frac{U^\mu U^\nu}{c^2}-pg^{\mu\nu}, \qquad \hbox{where} \qquad \epsilon=\rho c^2\left(1+\frac\varepsilon{c^2}\right).
\een
Here $U^\mu$ with $U^\mu U_\mu=c^2$ denotes the fluid four-velocity. The energy density has two contributions, one connected with the mass density $\rho=mn$ ($m$ is the rest mass of a fluid particle and $n$ the particle number density of the fluid),  and the other with the internal energy density $\rho\varepsilon$. In the works \cite{Ch0,Ch1,Ch2,Ch3,Nu,Ch4} the specific internal energy $\varepsilon$ was denoted by $\Pi$.

\begin{table}[h]
    \centering
    \begin{tabular}{c|ccc}\hline
       Equations   & Orders of &the metric &coefficients needed  \\
        of motion & $g_{00}$&$g_{0i}$&$g_{ij}$\\\hline
         Newtonian approximation&$1-2U/c^2$&0&$-\delta_{ij}$\\
         $\frac12$-post-Newtonian&0&0&0\\
         1-post-Newtonian&$2(U^2-2\Phi)/c^{-4}$&$\Pi_i/c^{-3}$&$-2U\delta_{ij}/c^{-2}$\\
         $1\frac12$-post-Newtonian&${\Psi^{(5)}_{00}}/{c^5}$&0&0\\
         $2$-post-Newtonian&${\Psi^{(6)}_{00}}/{c^6}$&${\Psi^{(5)}_{0i}}/{c^5}$&${\Psi^{(4)}_{ij}}/{c^4}$\\
         $2\frac12$-post-Newtonian&${\Psi^{(7)}_{00}}/{c^7}$&${\Psi^{(6)}_{0i}}/{c^6}$&${\Psi^{(5)}_{ij}}/{c^5}$\\\hline
    \end{tabular}
    \caption{Metric tensor coefficients necessary in the various approximations by reference to \cite{Ch4,Ch5}.  The correspondence of the gravitational potentials given here with those in  \cite{Ch4,Ch5} are: $\Pi_i\rightarrow P_i$, $\Psi^{(n)}_{ij}\rightarrow Q^{(n)}_{ij}$, $\Psi^{(n)}_{0i}\rightarrow Q^{(n)}_{0i}$ and $\Psi^{(n)}_{00}\rightarrow Q^{(n)}_{00}$.}
    \label{tab1}
\end{table}

In  Table \ref{tab1}  the metric tensor coefficients in the various approximations are given, according to \cite{Ch4,Ch5}. The expressions for the covariant and contravariant metric tensor components in the different post-Newtonian approximations, which are solutions of Einstein's field equations (\ref{ON1}) and obey the de Donder (harmonic) coordinate conditions ${\mathfrak{g}^{\mu\nu}}_{,\nu}=0$ are (see \cite{Ch4,Ch5})
\ben\lb{sp1a}
&&g_{00}=1-\frac{2U}{c^2}+\frac2{c^4}\left(U^2-2\Phi\right)+ \frac{\Psi^{(5)}_{00}}{c^5}+ \frac{\Psi^{(6)}_{00}}{c^6}+\frac{\Psi^{(7)}_{00}}{c^7},
\\\lb{sp1b}
&&g_{0i}=\frac{\Pi_i}{c^3}+\frac{\Psi^{(5)}_{0i}}{c^5}+\frac{\Psi^{(6)}_{0i}}{c^6},
\\\lb{sp1c}
&&g_{ij}=-\left(1+\frac{2U}{c^2}\right)\delta_{ij}+\frac{\Psi^{(4)}_{ij}}{c^4}+\frac{\Psi^{(5)}_{ij}}{c^5},
\\\lb{sp1a1}
&&g^{00}=1+\frac{2U}{c^2}+\frac2{c^4}\left(U^2+2\Phi\right)- \frac{\Psi^{(5)}_{00}}{c^5}- \frac1{c^6}\left({\Psi^{(6)}_{00}}+\Pi_k^2-16U\Phi\right)-\frac1{c^7}\left({\Psi^{(7)}_{00}}+4U{\Psi^{(5)}_{00}}\right),
\\\lb{sp1b1}
&&g^{0i}=\frac{\Pi_i}{c^3}+\frac{\Psi^{(5)}_{0i}}{c^5}+\frac{\Psi^{(6)}_{0i}}{c^6},
\\\lb{sp1c1}
&&g^{ij}=-\left(1-\frac{2U}{c^2}\right)\delta_{ij}-\frac1{c^4}\left({\Psi^{(4)}_{ij}}+4U^2\delta_{ij}\right)-\frac{\Psi^{(5)}_{ij}}{c^5}.
\een

The components of the Christoffel symbols related with the components of the metric tensor   are given in Appendix A.

The Newtonian gravitational potential $U$ and the post-Newtonian gravitational potentials $\Phi$, $\Pi_i$, $\Psi^{(4)}_{ij},$ $\Psi^{(5)}_{0i}$, $\Psi^{(6)}_{00}$ and $\Psi^{(7)}_{00}$ that follow from the solution of Einstein's field equations   are ruled by the Poisson equations (see \cite{Ch2,Ch4})
\ben\lb{sp2a}
&&\nabla^2U=-4\pi G\rho,\qquad\nabla^2\Phi=-4\pi G\rho\left(V^2+U+\frac\varepsilon2+\frac{3p}{2\rho}\right),
\qquad
\nabla^2\Pi_i=-16\pi G\rho V_i+\frac{\partial^2U}{\partial t\partial x^i},
\\\no
&&\nabla^2\Psi^{(4)}_{ij}=16\pi G\rho\bigg(V_iV_j-V^2\delta_{ij}-2\frac{p}\rho\delta_{ij}\bigg)-2\bigg(\delta_{ij}\nabla^2+\frac{\partial^2}{\partial x^i\partial x^j}\bigg)(U^2+2\Phi)
+4\frac{\partial U}{\partial x^i}\frac{\partial U}{\partial x^j}-2\frac{\partial^2U}{\partial t^2}\delta_{ij}
\\\lb{sp2b}
&&\qquad-\frac{\partial}{\partial t}\bigg(\frac{\partial \Pi_i}{\partial x^j}+\frac{\partial \Pi_j}{\partial x^i}\bigg),
\\\lb{sp2c}
&&\nabla^2\Psi^{(5)}_{0i}=-16\pi G\rho\left[V_i\left(V^2+\varepsilon+\frac{p}\rho+4U\right)-\frac{\Pi_i}2\right]-10\frac{\partial U}{\partial t}\frac{\partial U}{\partial x^i}
-2\frac{\partial U}{\partial x^j}\frac{\partial \Pi_j}{\partial x^i}+2\Pi_j\frac{\partial^2 U}{\partial x^i\partial x^j},
\\\no
&&\nabla^2\Psi^{(6)}_{00}=16\pi G\rho\left[V^2\left(V^2+\varepsilon+\frac{p}\rho+4U\right)-U^2-2\Phi\right]
+2\frac{\partial U}{\partial x^i}\frac{\partial \Pi_i}{\partial t}-6\left(\frac{\partial U}{\partial t}\right)^2+12\frac{\partial U}{\partial x^i}\frac{\partial \Phi}{\partial x^i}
\\\lb{sp2d}
&&\qquad-12U\left(\frac{\partial U}{\partial x^i}\right)^2+\frac{\partial \Pi_j}{\partial x^i}\left(\frac{\partial \Pi_i}{\partial x^j}-\frac{\partial \Pi_j}{\partial x^i}\right)+2\Psi_{ij}\frac{\partial^2 U}{\partial x^i\partial x^j},
\\\lb{sp2e}
&&\nabla^2\Psi^{(7)}_{00}=8 \pi G\rho \Psi^{(5)}_{00}+2\Psi^{(5)}_{ij}\frac{\partial^2U}{\partial x^i\partial x^j}+\frac{\partial^2\Psi^{(5)}_{00}}{\partial t^2}.
\een
Here  $V_i$ is the hydrodynamic velocity of the fluid. Furthermore, for the gravitational potentials $\Psi^{(5)}_{00},$ $\Psi^{(5)}_{ij}$ and $\Psi^{(6)}_{0i}$  the following equations hold (see \cite{Ch5})
\ben\lb{q1}
\Psi^{(5)}_{00}=\frac43G\frac{d^3I_{kk}}{dt^3},
\qquad
\Psi^{(5)}_{ij}=2G\frac{d^3I_{ij}}{dt^3}-\frac23G\delta_{ij}\frac{d^3I_{kk}}{dt^3},\qquad \Psi^{(5)}_{kk}=0,
\\\lb{q2}
\Psi^{(6)}_{0i}=\frac23Gx_j\frac{d^4I_{ij}}{dt^4}-\frac23G\frac{d^3}{dt^3}\int\rho V_i\vert\bx\vert^2d\bx,\qquad \frac{d\Psi^{(5)}_{00}}{dt}=2\frac{\partial \Psi^{(6)}_{0i}}{\partial x^i},\qquad \frac{\partial \Psi^{(6)}_{0i}}{\partial x^j}=\frac{\partial \Psi^{(6)}_{0j}}{\partial x^i}.
\een
where $I_{ij}=\int_V\rho x_ix_jd\bx$ denotes the moment of inertia tensor. %The gravitational potential $\Psi^{(7)}_{00}$ is ruled by the relationship
%\ben
%\Psi^{(7)}_{00}=-2\Psi^{(5)}_{00}U-\Psi^{(5)}_{ij}\frac{\partial^2\chi}{\partial x^i\partial x^j}+\frac1{60}G\frac{\partial^5}{\partial t^5}\int\rho(\bx',t)\vert\bx-\bx'\vert^4 d\bx'+\frac13\frac{\partial^3}{\partial t^3}\int\Theta_{ii}(\bx',t)\vert\bx-\bx'\vert^2 d\bx',\een
%where the super-potential $\chi$  obeys the equation $\nabla^4\chi=8\pi G\rho$ and $\Theta_{ij}$ is the Newtonian expression for the energy-momentum complex:
%\ben\Theta_{ij}=\rho V_i V_j+p\delta_{ij}+\frac1{16\pi G}\left[4\frac{\partial U}{\partial x^i}\frac{\partial U}{\partial x^j}-2\delta_{ij}\frac{\partial U}{\partial x^k}\frac{\partial U}{\partial x^k}\right].\een

 The  four-velocity hydrodynamic components are given by
\ben\lb{sp3a}
(U^\mu)=\left(\frac{dx^0}{d\tau}, \frac{dx^i}{d\tau}\right)=\left(c\frac{dt}{d\tau},V_i\frac{dt}{d\tau}\right),
\een
and are obtained from  expression that follows from the line element   
\ben\lb{sp3b}
\left(\frac{d\tau}{dt}\right)^2=g_{00}+\frac2cg_{0i}V^i+\frac1{c^2}g_{ij}V^iV^j,
\een
together with  the components of the metric tensor (\ref{sp1a}) -- (\ref{sp1c}), yielding the following expression up to the order $1/c^7$:
\ben\no
&&U^0=c\left\{1+\frac1{c^2}\left(\frac{V^2}2+U\right)
+\frac1{c^4}\left(\frac{3V^4}8+\frac{5U V^2}2+\frac{U^2}2+2\Phi-\Pi_iV_i\right)-\frac{\Psi^{(5)}_{00}}{2c^5}\right.
\\\no
&&\qquad\left.+\frac1{c^6}\left[\frac5{16}V^6+\frac{27}8 UV^4
+\frac{21}4 U^2V^2-\frac{U^3}2+3\left(V^2+2U\right)\left(\Phi-\frac{\Pi_iV_i}{2}\right)-\frac12\left(\Psi^{(6)}_{00}+V_iV_j\Psi^{(4)}_{ij}+2V_i\Psi^{(5)}_{0i}\right)\right]\right.
\\\lb{sp4}
&&\qquad\left.-\frac1{2c^7}\left[\frac32\left(V^2+2U\right)\Psi^{(5)}_{00}+V_iV_j\Psi^{(5)}_{ij}+2\Psi^{(6)}_{0i}V_i+\Psi^{(7)}_{00}\right]\right\},
\qquad
U^i=\frac{V_iU^0}c.
\een

\section{Boltzmann Equation in $2\frac12$--PN}
\lb{s.3}

A  gas  with rest mass particles $m$ is described in the relativistic kinetic theory  by its space-time coordinates $(x^\mu)=(ct,\bx)$ and the momentum four-vector $(p^\mu)=(p^0,\bp)$.  The length of the momentum four-vector is constant and constrained by the relationship $g_{\mu\nu}p^\mu p^\nu=m^2c^2$.

In the phase space spanned by the  spatial coordinates $\bx$ and the momentum $\bp$ of the particles, the state of the gas is characterized by the  invariant one-particle distribution function $f(\bx,\bp,t)$, such that the number of particle world lines that cross the hypersurface element represented by the three-dimensional space $d^3X=p^0\sqrt{-g}d^3x$ on the surface $x^0=$ constant and with the  four-vector spatial momentum contained in the cell $d^3P=\sqrt{-g}\,{d^3p}/{p_0}$ of the mass-shell,
is the invariant $dN=f(\bx,\bp,t)(-g)d^3xd^3p$ \cite{CK}.

The space-time evolution of the one-particle distribution function $f(\bx,\bp,t)$ is ruled by the Boltzmann equation, which is a non-linear integro-differential equation. The first and second post-Newtonian approximations of the Boltzmann equation   were   derived in the papers \cite{Rez,Ped,GK2}. 

 Let $v^i=c p^i/p^0$ be the  three-velocity of a particle  and $\tau$  be the proper time along the world line of the particle. We represent  the one-particle distribution function as $f(x^\mu(\tau),v_i(\tau))$, so that  its variation  with respect to the proper time can be written as
\ben\lb{k3}
\frac{df(x^\mu(\tau),v_i(\tau))}{d\tau}=\frac{\partial f}{\partial x^\mu}\frac{dx^\mu}{d\tau}+\frac{\partial f}{\partial v_i}\frac{dv_i}{d\tau}=\frac{u^0}c\left(\frac{\partial f}{\partial t}+v_i\frac{\partial f}{\partial x^i}+\frac{\partial f}{\partial v_i}\frac{d^2x^i}{dt^2}\right).
\een
Here $(u^\mu)=(u^0,u^0v_i/c)$ is the four-velocity of the gas particles, whose components have expressions  similar to those given in (\ref{sp4}), once the hydrodynamic velocity $\bV$ is replaced  by the particle velocity $\bv$.

From the equation of motion of the gas particles \cite{Wein}
\ben\lb{k5}
\frac{d^2x^\mu}{d\tau^2}+{\Gamma^\mu}_{\nu\lambda}\frac{dx^\nu}{d\tau}\frac{dx^\lambda}{d\tau}=0,
\een
we can determine the acceleration, which can be written as
\ben\no
&&\frac{d^2 x^i}{d(x^0)^2}=\left(\frac{dx^0}{d\tau}\right)^{-2}\frac{dx^\mu}{d\tau}
\frac{dx^\nu}{d\tau}\bigg[{\Gamma^0}_{\mu\nu}\left(\frac{dx^0}{d\tau}\right)^{-1}\frac{dx^i}{d\tau}
-{\Gamma^i}_{\mu\nu}\bigg]
=-{\Gamma^i}_{00}-{\Gamma^i}_{jk}\frac{dx^j}{dx^0}\frac{dx^k}{dx^0}-2{\Gamma^i}_{0j}\frac{dx^j}{dx^0}
\\\lb{k7}
&&\qquad+\frac{dx^i}{dx^0}\left[{\Gamma^0}_{00}+2{\Gamma^0}_{0j}\frac{dx^j}{dx^0}
+{\Gamma^0}_{jk}\frac{dx^j}{dx^0}\frac{dx^k}{dx^0}\right].\,
\een

The Boltzmann equation in the $2\frac12$--post--Newtonian approximation follows from (\ref{k3}) together with (\ref{k7}) -- which represents the streaming operator ${\mathcal{L}}_{2.5PN}[f]$ --  and the collision operator $\mathcal{C}[f]$, namely ${\mathcal{L}}_{2.5PN}[f]=\mathcal{C}[f]$.

Here we are interested in deriving  from the Boltzmann equation the Euler equations which correspond to the hydrodynamic equations of mass, mass-energy and momentum densities. These quantities are defined in terms of the collision invariants of mass and momentum four-vector of a gas particle, which implies a vanishing collision operator. Hence, below, we shall introduce the collisionless Boltzmann equation.  

If we insert the expressions for the Christoffel symbol components in Appendix A  into (\ref{k7}) and combine with (\ref{k3}), we get -- up to order $1/c^{7}$ -- the space-time evolution equation for the one-particle distribution function, which represents  the  collisionless Boltzmann equation in the $2\frac12$-post-Newtonian approximation, namely
\begin{small}
\ben\no
&&\frac{\partial f}{\partial t}+v_i\frac{\partial f}{\partial x^i}+\frac{\partial f}{\partial v_i}\frac{\partial U}{\partial x^i}+\frac{\partial f}{\partial v_i}\Bigg\{v_i\bigg\{-\frac1{c^2}\frac{\partial U}{\partial t}+\frac1{c^3}\left(\Pi_i\frac{\partial U}{\partial x^i}-2\frac{\partial \Phi}{\partial t}\right)+\frac1{2c^5}\frac{d\Psi^{(5)}_{00}}{dt}+\frac1{c^7}\bigg[\frac12\frac{\partial\Psi^{(6)}_{00}}{\partial t}+2U^2\frac{\partial U}{\partial t}-4\bigg(U\frac{\partial \Phi}{\partial t}
\\\no
&&+\Phi\frac{\partial U}{\partial t}\bigg)+{\Pi_j}\frac{\partial \Pi_j}{\partial t}-\Pi_j\frac{\partial (U^2-2\Phi)}{\partial x^j}+\Psi^{(5)}_{0j}\frac{\partial U}{\partial x^j}\bigg]+\frac1{c^7}\left[\frac12\frac{\partial\Psi^{(7)}_{00}}{\partial t}+U\frac{\partial \Psi^{(5)}_{00}}{\partial t}+\Psi^{(5)}_{00}\frac{\partial U}{\partial t}+\Psi^{(6)}_{0i}\frac{\partial U}{\partial x^i}\right]\bigg\}
\\\no
&&+2v_iv_j\bigg\{-\frac1{c^2}\frac{\partial U}{\partial x^j}-\frac2{c^4}\frac{\partial\Phi}{\partial x^j}+\frac1{c^6}\left[\frac12\frac{\partial \Psi^{(6)}_{00}}{\partial x^j}+2U^2\frac{\partial U}{\partial x^j}-4\left(U\frac{\partial \Phi}{\partial x^j}+\Phi\frac{\partial U}{\partial x^j}\right)+\frac{\Pi_k}{2}\left(\frac{\partial \Pi_k}{\partial x^j}-\frac{\partial \Pi_j}{\partial x^k}-2\frac{\partial U}{\partial t}\delta_{jk}\right)\right]
\\\no
&&+\frac1{c^7}\left[\frac12\frac{\partial \Psi^{(7)}_{00}}{\partial x^j}+\Psi^{(5)}_{00}\frac{\partial U}{\partial x^j}\right]\bigg\}-\frac1{c^2}\left[2\frac{\partial (U^2-\Phi)}{\partial x^i}-\frac{\partial\Pi_i}{\partial t}\right]-\frac1{c^4}\left[\frac12\frac{\partial\Psi^{(6)}_{00}}{\partial x^i}-\frac{\partial\Psi^{(5)}_{0i}}{\partial t}-\Pi_i\frac{\partial U}{\partial t}-4U\bigg(\frac{\partial U^2}{\partial x^i}-\frac{\partial \Phi}{\partial x^i}\bigg)\right.
\\\no
&&
\left.+2U\frac{\partial \Pi_i}{\partial t}-\Psi^{(4)}_{ij}\frac{\partial U}{\partial x^j}\right]
-\frac1{2c^5}\left[\frac{\partial\Psi^{(7)}_{00}}{\partial x^i}-2\frac{\partial\Psi^{(6)}_{0i}}{\partial t}-2\Psi^{(5)}_{ij}\frac{\partial U}{\partial x^j}\right]-\frac1{c^6}\left[\Pi_i\frac{\partial(U^2-2\Phi)}{\partial t}-\Psi^{(5)}_{0i}\frac{\partial U}{\partial t}+U\left(2\frac{\partial\Psi^{(5)}_{0i}}{\partial t}-\frac{\partial\Psi^{(6)}_{00}}{\partial x^i}\right)\right.
\\\no
&&\left.-\left(\Psi^{(4)}_{ij}+4U^2\delta_{ij}\right)\left(\frac{\partial \Pi_j}{\partial t}-\frac{\partial(U^2-2\Phi)}{\partial x^j}\right)\right]-\frac1{2c^7}\left[\Pi_i\frac{\partial\Psi^{(5)}_{00}}{\partial t}+2U\left(2\frac{\partial\Psi^{(6)}_{0i}}{\partial t}-\frac{\partial\Psi^{(7)}_{00}}{\partial x^i}\right)-2\Psi^{(5)}_{ij}\left(\frac{\partial \Pi_j}{\partial t}-\frac{\partial(U^2-2\Phi)}{\partial x^j}\right)\right]
\\\no
&&+v_iv_jv_k\bigg\{\frac1{2c^4}\left(\frac{\partial\Pi_k}{\partial x^j}+\frac{\partial\Pi_j}{\partial x^k}+2\frac{\partial U}{\partial t}\delta_{kj}\right)+\frac1{c^6}\left[U\left(\frac{\partial \Pi_k}{\partial x^j}+\frac{\partial \Pi_j}{\partial x^k}+2\delta_{kj}\frac{\partial U}{\partial t}\right)-\left(\Pi_k\frac{\partial U}{\partial x^j}+\Pi_j\frac{\partial U}{\partial x^k}-\Pi_l\frac{\partial U}{\partial x^l}\delta_{kj}\right)\right.
\\\no
&&\left.+\frac12\left(\frac{\partial \Psi^{(5)}_{0k}}{\partial x^j}+\frac{\partial \Psi^{(5)}_{0j}}{\partial x^k}-\frac{\partial \Psi^{(4)}_{kj}}{\partial t}\right)\right]+\frac1{2c^7}\left[\frac{\partial \Psi^{(6)}_{0k}}{\partial x^j}+\frac{\partial \Psi^{(6)}_{0j}}{\partial x^k}-\frac{\partial \Psi^{(5)}_{kj}}{\partial t}\right]\bigg\}-2v_j\bigg\{\frac1{2c^2}\left(\frac{\partial\Pi_j}{\partial x^i}-\frac{\partial\Pi_i}{\partial x^j}+2\frac{\partial U}{\partial t}\delta_{ij}\right)
\\\no
&&-\frac1{c^4}\left[\Pi_i\frac{\partial U}{\partial x^j}-U\left(\frac{\partial\Pi_i}{\partial x^j}-\frac{\partial \Pi_j}{\partial x^i}-2\frac{\partial U}{\partial t}\delta_{ij}\right)+\frac1{2}\left(\frac{\partial \Psi^{(5)}_{0i}}{\partial x^j}-\frac{\partial \Psi^{(5)}_{0j}}{\partial x^i}+\frac{\partial\Psi^{(4)}_{ij}}{\partial t}\right)\right]-\frac1{2c^5}\frac{d\Psi^{(5)}_{ij}}{dt}+\frac1{c^6}\left[\Pi_i\frac{\partial (U^2-2\Phi)}{\partial x^j}\right.
\\\no
&&\left.-\frac12\left(\Psi^{(4)}_{ik}+4U^2\delta_{ik}\right)\left(\frac{\partial\Pi_k}{\partial x^j}-\frac{\partial \Pi_j}{\partial x^k}-2\frac{\partial U}{\partial t}\delta_{jk}\right)+U\left(\frac{\partial \Psi^{(5)}_{0i}}{\partial x^j}-\frac{\partial \Psi^{(5)}_{0j}}{\partial x^i}+\frac{\partial\Psi^{(4)}_{ij}}{\partial t}\right)-\Psi^{(5)}_{0i}\frac{\partial U}{\partial x^j}\right]+\frac1{c^7}\left[U\frac{\partial \Psi^{(5)}_{ij}}{\partial t}\right.
\\\no
&&\left.-\Psi^{(6)}_{0i}\frac{\partial U}{\partial x^j}-\frac{\Psi^{(5)}_{ik}}2\left(\frac{\partial\Pi_k}{\partial x^j}-\frac{\partial \Pi_j}{\partial x^k}-2\frac{\partial U}{\partial t}\delta_{jk}\right)\right]\bigg\}-v_iv_jv_k\bigg\{\frac1{c^2}\left(\frac{\partial U}{\partial x^j}\delta_{ik}+\frac{\partial U}{\partial x^k}\delta_{ij}-\frac{\partial U}{\partial x^i}\delta_{jk}\right)-\frac1{c^4}\left[\frac12\left(\frac{\partial\Psi^{(4)}_{ij}}{\partial x^k}\right.\right.
\\\no
&&\left.\left.+\frac{\partial\Psi^{(4)}_{ik}}{\partial x^j}-\frac{\partial\Psi^{(4)}_{jk}}{\partial x^i}\right)
+\left(\frac{\partial U^2}{\partial x^k}\delta_{ij}+\frac{\partial U^2}{\partial x^j}\delta_{ik}-\frac{\partial U^2}{\partial x^i}\delta_{jk}\right)\right]+\frac1{c^6}\left[\left(\Psi^{(4)}_{il}+4U\delta_{il}\right)\left(\frac{\partial U}{\partial x^k}\delta_{jl}+\frac{\partial U}{\partial x^j}\delta_{kl}-\frac{\partial U}{\partial x^l}\delta_{jk}\right)\right.
\\\no
&&\left.+\frac{\Pi_i}{2}\left(\frac{\partial\Pi_k}{\partial x^j}+\frac{\partial\Pi_j}{\partial x^k}+2\frac{\partial U}{\partial t}\delta_{jk}\right)+U\left(\frac{\partial\Psi^{(4)}_{ij}}{\partial x^k}+\frac{\partial\Psi^{(4)}_{ik}}{\partial x^j}-\frac{\partial\Psi^{(4)}_{jk}}{\partial x^i}\right)\right]+\frac1{c^7}\left[\frac{\partial U\Psi^{(5)}_{ij}}{\partial x^k}+\frac{\partial U\Psi^{(5)}_{ik}}{\partial x^j}-\frac{\partial U\Psi^{(5)}_{jk}}{\partial x^i}\right.
\\\lb{bol}
&&\left.+\frac{\partial U}{\partial x^l}\left(\Psi^{(5)}_{jk}\delta_{li}-\Psi^{(5)}_{il}\delta_{jk}\right)\right]\bigg\}\Bigg\}=0.
\een
\end{small}

\section{Maxwell--J\"uttner Distribution Function in $\mathbf{2\frac12}$--PN}\lb{s.4}

The  Maxwell--J\"uttner distribution function is a relativistic expression for the one-particle distribution function at equilibrium. It follows from the collision term of the relativistic Boltzmann equation, by requiring that it vanishes at equilibrium, once the number of particles that enter and leave the volume element  in the phase space are the same (see \cite{CK}). The Maxwell--J\"uttner distribution function is given by 
\ben\lb{mj1}
f(\bx,\bp,t)=\frac{n}{4\pi m^2ckTK_2(\zeta)}\exp\left(-\frac{g_{\mu\nu}p^\mu U^\nu}{kT}\right).
\een
In the above equation $T$ is the absolute temperature and $k$ the Boltzmann constant. Furthermore,  $K_2(\zeta)$ represents the modified Bessel function of the second kind, which depends on the relativistic parameter $\zeta=mc^2/kT$,  the ratio of the rest energy of the gas particles $mc^2$ and the thermal energy of the gas $kT$. Two limiting cases are: (i)  $\zeta\gg1$ in the non-relativistic limit, (ii) $\zeta\ll1$   in the ultra-relativistic limit.

The Maxwell--J\"uttner distribution function in the desired post-Newtonian approximation follows from the expansion of the exponential term in powers up to $1/c^7$, once the term  $g_{\mu\nu}U^\mu u^\nu$ is evaluated up to $1/c^7$ order terms  and by considering the peculiar velocity $W_i=v_i-V_i$,  which is the difference of the particle velocity and the gas velocity, i.e. it refers to the particle velocity  in the gas frame (see \cite{KRW,GK1,GK2} for more details). It is also necessary  to 
evaluate up to  $1/c^6$ terms the modified Bessel function of second kind, which reads \cite{AbSt}
\ben\lb{mj3}
\frac1{K_2(\zeta)}=\sqrt{\frac{2mc^2}{\pi kT}}\,e^{\frac{mc^2}{kT}}
\left(1-\frac{15kT}{8mc^2}+\frac{345(kT)^2}{128m^2c^4}-\frac{3285(kT)^3}{1024m^3c^6} \right).
\een

Taking into account the method described above, the Maxwell--J\"uttner distribution function up to $1/c^7$ order terms, becomes
\ben\nonumber
&&f=\frac{n}{(2\pi mkT)^{\frac32}}e^{-\frac{mW^2}{2kT}}
\Bigg\{ 1-\frac{1}{c^2}\bigg[\frac{15 kT}{8 m}+\frac{m (VW)^2}{2 kT}+\frac{2 m U W^2}{kT}+\frac{3 m W^4}{8 kT}+\frac{m V^2 W^2}{2 kT}+\frac{m (V W) W^2}{kT}\bigg]
\\\no
&&+\frac{1}{c^4}\bigg[\frac{2 m^2 U^2 W^4}{(kT)^2}+\frac{m^2 U V^2 W^4}{(kT)^2}+\frac{3 m^2 U W^6}{4 (kT)^2}
+\frac{m^2 U (V W)^2 W^2}{(kT)^2}+\frac{2 m^2 U (V W) W^4}{(kT)^2}+\frac{3 m^2 V^2 W^6}{16 (kT)^2}+\frac{m^2 V^4 W^4}{8 (kT)^2}
\\\no
&&+\frac{m^2 (V W)^4}{8 (kT)^2}+\frac{m^2 V^2 (V W)^2 W^2}{4 (kT)^2}+\frac{m^2 V^2 (V W) W^4}{2 (kT)^2}+\frac{m^2 (V W)^3 W^2}{2 (kT)^2}+\frac{11 m^2 (V W)^2 W^4}{16 (kT)^2}+\frac{3 m^2 (V W) W^6}{8 (kT)^2}
\\\no
&&+\frac{9 m^2 W^8}{128 (kT)^2}+\frac{345 (kT)^2}{128 m^2}
-\frac{3 m U^2 W^2}{kT}-\frac{4 m U V^2 W^2}{kT}-\frac{4 m U (V W)^2}{kT}-\frac{8 m U (V W) W^2}{kT}-\frac{3 m U W^4}{kT}
\\\no
&&+\frac{m \Pi_iW_i (V W)}{kT}-\frac{m V^4 W^2}{2 kT}-\frac{m V^2 (V W)^2}{kT}
-\frac{2 m V^2 (V W) W^2}{kT}-\frac{3 m V^2 W^4}{4 kT}-\frac{m (V W)^3}{kT}-\frac{9 m (V W)^2 W^2}{4 kT}
\\\no
&&-\frac{3 m (V W) W^4}{2 kT}-\frac{5 m W^6}{16 kT}+\frac{m (\Pi V) W^2}{kT}
+\frac{m (\Pi W) W^2}{kT}-\frac{2 m \Phi  W^2}{kT}+\frac{m W_i W_j \Psi^{(4)}_{ij}}{2 kT}+\frac{15 U W^2}{4}+\frac{15 V^2 W^2}{16}
\\\lb{mj4}
&&+\frac{15 (V W)^2}{16}+\frac{15 (V W) W^2}{8}+\frac{45 W^4}{64}\bigg]+\frac1{c^5}\frac{m}{2kT}\left[W^2\Psi^{(5)}_{00}+W_iW_j\Psi^{(5)}_{ij}\right]+\frac{f^{(6)}}{c^6}+\frac{f^{(7)}}{c^7}\bigg\}.
\een
Here we have introduced the abbreviations $VW\equiv V_k W_k$, $\Pi V\equiv \Pi_k V_k$,  $\Pi W\equiv \Pi_k W_k$. Furthermore, the expressions corresponding to the approximations $1/c^6$ and $1/c^7$, denoted by $f^{(6)}$ and $f^{(7)}$, can be found in  Appendix B.

\section{Energy-momentum tensor in $2\frac12$--PN}\lb{s.5}

We can use the  Maxwell--J\"uttner distribution function (\ref{mj4}) to determine the components of the particle four-flow $N^\mu$ and of the energy-momentum tensor $T^{\mu\nu}$, since they are defined in the kinetic theory of relativistic gases as (see e.g. \cite{CK})
\ben\lb{mj5}
N^\mu=m^3c\int u^\mu  f\frac{\sqrt{-g}\,d^3 u}{u_0},
\qquad T^{\mu\nu}=m^4c\int u^\mu u^\nu f\frac{\sqrt{-g}\,d^3 u}{u_0}.
\een 

We note from (\ref{mj5}) that both expressions  are functions of the invariant element $\sqrt{-g}\,d^3 u/{u_0}$.  For more details on the transformation $d^3u\rightarrow d^3v$ one is referred to the works \cite{KRW,GK1,GK2}. Here we give only the final result:
\ben\no
&&\frac{\sqrt{-g}\, d^3 u}{u_0}=
\bigg\{1+\frac1{c^2}\bigg[\frac{5v^2}{2}
+7U\bigg]+\frac1{c^4}\bigg[\frac{35v^4}8+\frac{55Uv^2}2
 +\frac{43U^2}2+8\Phi-\frac{\Psi^{(4)}_{kk}}2-5\Pi_iv_i\bigg]-\frac{2}{c^5}\bigg[\Psi^{(5)}_{00}+\frac{\Psi^{(5)}_{kk}}4\bigg]
\\\no\lb{mj8}
&&\qquad+\frac1{c^6}\bigg[\frac{105v^6}{16}+\frac{525Uv^4}8+\frac{555U^2v^2}4+\frac{75U^3}2+\frac{\Pi_k^2}2-\bigg(\frac{35v^2}2+45U\bigg)\Pi_iv_i+\bigg(30v^2+72U\bigg)\Phi
\\\no
&&\qquad-\frac52v_iv_j\Psi^{(4)}_{ij}-\frac52\bigg(\frac{v^2}2+U\bigg)\Psi^{(4)}_{kk}-5v_i\Psi^{(5)}_{0i}-2\Psi^{(6)}_{00}\bigg]-\frac1{c^7}\bigg[\bigg(\frac{15v^2}2+18U\bigg)\Psi^{(5)}_{00}+\frac52v_iv_j\Psi^{(5)}_{ij}
\\\lb{inv}
&&\qquad+\frac52\bigg(\frac{v^2}2+U\bigg)\Psi^{(5)}_{kk}+5v_i\Psi^{(6)}_{0i}+2\Psi^{(7)}_{00}\bigg]\bigg\}\frac{d^3v}{u^0}.
\een

To perform the  integrations, we insert  the  Maxwell--J\"uttner distribution function (\ref{mj4}) and the integration element (\ref{mj8}) into (\ref{mj5}),  introduce the peculiar velocity $W_i=v_i-V_i$, the  spherical coordinates $(W,\theta,\varphi)$ and  express the integral element as $d^3W=W^2\sin\theta d\theta d\varphi dW$, where $0\leq W<\infty$, $0\leq\theta\leq\pi$ and $0\leq\varphi\leq2\pi$.  For the integrations, it is necessary to use the table of integrals given in the works \cite{GK1,GK2,GK3,GK9}.  

As should be, the particle four-flow is proportional to the four-velocity, i.e. $N^\mu=nU^\mu$, while the components of the energy momentum tensor read
\ben\no\lb{mj9a}
&&T^{00}=\rho c^2\bigg\{1+\frac1{c^2}\bigg(V^2+2U+\varepsilon\bigg)+\frac1{c^4}\bigg(V^4+6UV^2+2U^2
+\bigg(\varepsilon+\frac{p}\rho\bigg)V^2
+2\varepsilon U-2\Pi_jV_j+4\Phi\bigg)-\frac{\Psi^{(5)}_{00}}{c^5}
\\\no
&&\qquad+\frac1{c^6}\bigg[V^6+2V^2U(5V^2+8U)+8(V^2+2U)\Phi+2(U^2+2\Phi)\varepsilon+(V^4+6V^2U-2V_j\Pi_j)\left(\varepsilon+\frac{p}{\rho}\right)-\Psi^{(6)}_{00}
\\
&&\qquad-4(V^2+2U)V_j\Pi_j-2\Psi^{(5)}_{0j}V_j-\Psi^{(4)}_{jk}V_jV_k\bigg]-\frac1{c^7}\bigg[\Psi^{(7)}_{00}+2\Psi^{(6)}_{0j}V_j+\Psi^{(5)}_{jk}V_jV_k+\Psi^{(5)}_{00}(2V^2+4U+\varepsilon)\bigg]\bigg\},
\\\no\lb{mmj9c}
&&T^{0i}=\rho cV_i\bigg\{1+\frac1{c^2}\bigg(V^2+2U+\varepsilon+\frac{p}\rho\bigg)+\frac1{c^4}\bigg(V^4+6UV^2+2U^2
+\bigg(\varepsilon+\frac{p}\rho\bigg)(V^2+2 U)-2\Pi_jV_j+4\Phi\bigg)
\\\no
&&\qquad-\frac{\Psi^{(5)}_{00}}{c^5}+\frac1{c^6}\bigg[V^6+2V^2U(5V^2+8U)+8(V^2+2U)\Phi+(V^4+6V^2U-2V_j\Pi_j+2U^2+4\Phi)\left(\varepsilon+\frac{p}{\rho}\right)
\\\no
&&\qquad-\Psi^{(6)}_{00}-4(V^2+2U)V_j\Pi_j-2\Psi^{(5)}_{0j}V_j-\Psi^{(4)}_{jk}V_jV_k\bigg]%-\frac1{c^7}\bigg[\Psi^{(7)}_{00}+2\Psi^{(6)}_{0j}V_j+\Psi^{(5)}_{jk}V_jV_k%
%\\&&\qquad+\Psi^{(5)}_{00}\bigg(2V^2+4U+\varepsilon+\frac{p}\rho\bigg)\bigg]%
\bigg\}-\frac{p\Pi_i}{c^3}-\frac{p\Psi^{(5)}_{0i}}{c^5}%-\frac{p\Psi^{(6)}_{0i}}{c^6}%,
\\\no\lb{mmj9d}
&&T^{ij}=\rho\left(V_iV_j+\frac{p}\rho\delta_{ij}\right)+\frac\rho{c^2}\bigg[\bigg(V^2+2U
+\varepsilon +\frac{p}\rho\bigg)V_iV_j -\frac{2pU}\rho\delta_{ij}\bigg]
+\frac\rho{c^4}\bigg[\bigg(V^4+6UV^2+2U^2
\\
&&\qquad+\left(\varepsilon+\frac{p}\rho\right)(V^2+2U)-2\Pi_kV_k+4\Phi\bigg)V_iV_j+\frac{p}{\rho}\left(4U^2\delta_{ij}+\Psi^{(4)}_{ij}\right)
\bigg]+\frac1{c^5}\left(p\Psi^{(5)}_{ij}-\rho V_i V_j \Psi^{(5)}_{00}\right).
\een
In the final expressions for the components of the particle four-flow and energy-momentum tensor, we have used the following representations for the pressure and specific internal energy:
\ben\lb{mmj9b}
 p=\frac{\rho kT}m,\qquad \varepsilon=\frac{3kT}{2m}\left(1+\frac{5kT}{4mc^2}-\frac{5(kT)^2}{4m^2c^4}+\frac{45(kT)^3}{64m^3c^6}\right).
\een

The  components of the energy-momentum tensor (\ref{mj9a}) -- (\ref{mmj9d})  match those  given in \cite{Ch2,Ch4} corresponding to the $2$-- and $2\frac12$-- post-Newtonian approximations.

\section{ Eulerian Hydrodynamic Equations in $2\frac12$ -- PN}\lb{s.6}

The Euler hydrodynamic equations refer to the balance equations of non viscous and non heat conducting fluids, which are described by its pressure and internal energy. Within the framework of kinetic theory of gases, the hydrodynamic equations are obtained from the Boltzmann equation, while in the phenomenological theory they follow from   the conservation equations of the particle four-flow $N^\mu$ and energy-momentum tensor $T^{\mu\nu}$,  namely
\ben\lb{ba1}
{N^\mu}_{;\mu}=\frac{\partial N^\mu}{\partial x^\mu}+{\Gamma^\mu}_{\mu\lambda}N^\lambda=0,
\qquad
{T^{\mu\nu}}_{;\nu}=\frac{\partial T^{\mu\nu}}{\partial x^\nu}+{\Gamma^\mu}_{\nu\lambda}T^{\lambda\nu}+{\Gamma^\nu}_{\nu\lambda}T^{\mu\lambda}=0.
\een

Here we shall follow the methodology of the kinetic theory of gases to obtain the hydrodynamic equations from the Boltzmann equation. From the local equilibrium assumption, we consider  $f$ as  the Maxwell--J\"uttner  distribution function (\ref{mj4}), and  multiply the Boltzmann equation (\ref{bol}) by the invariant element (\ref{inv}) and by an arbitrary function $\Upsilon$ of the  four-velocity of the gas particle. The hydrodynamic equations are obtained by considering a specific value for the arbitrary function $\Upsilon$ and by integrating  the resulting equation over all values of the  peculiar velocity $0\leq W<\infty$.

For the determination of the  hydrodynamic mass density equation, which is the continuity equation, one chooses $\Upsilon=m^4$, yielding 
\ben\lb{bal1a}
 \frac{\partial\widetilde\rho}{\partial t}+\frac{\partial \widetilde \rho V_i}{\partial x^i}=0,
 \een
 where the mass density $\widetilde\rho$ is given by
 \ben\no
 &&\widetilde\rho=\rho\left[1+\frac1{c^2}\left(\frac{V^2}2+3U\right)+\frac1{c^4}\left(\frac38V^4+\frac72UV^2+\frac32U^2-\frac12\Psi^{(4)}_{kk}-\Pi_iV_i\right)+\frac1{c^6}\left(\frac5{16}V^6+\frac{33}8V^4U+\frac{39}4V^2U^2\right.\right.
\\\no
&&\qquad\left.\left.-\frac12U^3+2V^2\Phi-\frac12(3V^2+10U)\Pi_kV_k+\frac12\Pi_k^2-\frac14(V^2+2U)\Psi^{(4)}_{kk}-V_i\Psi^{(5)}_{0i}-\frac12V_iV_j\Psi^{(4)}_{ij}\right)\right.
\\\lb{bal1b}
&&\qquad\left.-\frac\rho{2c^7}\left(V_iV_j\Psi^{(5)}_{ij}+2\Psi^{(6)}_{0i}V_i+\Psi^{(5)}_{00}V^2\right)\right].
 \een
The above expression for the mass density $\widetilde\rho$ corresponds to eqs. \emph{(95)} and \emph{(128)} of  the works \cite{Ch2,Ch4}, respectively, which was determined  considering that the volume integral of $\rho U^0\sqrt{-g}$ is constant,  a consequence of the particle four-flow conservation equation. The difference in the expression given here and the expressions quoted above is that they have the terms  $Q^{(3)}_{kk}/c^6$ and $Q^{(7)}_{kk}/c^7$ ($\Psi^{(3)}_{kk}/c^6$ and $\Psi^{(7)}_{kk}/c^7$ in our notation), which are contributions from the $3$ and $3 \frac12$ approximations to the components of the metric tensor $g_{ij}$, respectively. However, the expression for the mass density $\widetilde\rho$  given in (\ref{bal1b}) is consistent with the approximation of the Maxwell--J\"uttner distribution function (\ref{mj4}) and is a function of all gravitational potentials up to the $2\frac12$--post--Newtonian approximation.
The determination of the gravitational potential $Q^{(3)}_{kk}/c^6$ can be found in \cite{Ch2} pp. 75--77,  while $Q^{(7)}_{kk}/c^7$  in \cite{Ch4} pp. 175--176. Their expressions are not given here, because they are too long.

The hydrodynamic equation for the time component of the energy-momentum tensor conservation is the hydrodynamic mass-energy density equation. It is obtained by choosing $\Upsilon=m^4 u^0$, resulting in the expression
\ben\lb{bal2a}
&&\frac{\partial\widetilde\sigma}{\partial t}+\frac{\partial \widetilde\sigma V_i}{\partial x^i}+\Sigma=0,
\een
where the mass-energy density $\widetilde\sigma$ is given by
\ben\no\lb{bal2b}
&&\widetilde\sigma=\rho c^2\bigg\{1+\frac1{c^2}\bigg(V^2+3U+\varepsilon\bigg)+\frac1{c^4}\bigg[V^4+6V^2U-U^2
+2U\varepsilon+V^2\bigg(\varepsilon+\frac{p}\rho\bigg)-\Pi_iV_i-\frac12\Psi^{(4)}_{kk}\bigg]
\\\no
&&\qquad+\frac1{c^6}\bigg[V^6+\bigg(10U+\varepsilon+\frac{p}\rho\bigg)V^4+\bigg(13U+6\varepsilon+6\frac{p}\rho\bigg)UV^2-U^3 -\bigg(3V^2+4U+\varepsilon+\frac{p}\rho\bigg)\Pi_kV_k
\\\no
&&\qquad+\big(6V^2+4\varepsilon\big)\Phi+\frac12\Pi_k^2-\frac{\Psi^{(4)}_{kk}}{2}\bigg(\frac{V^2}2+U\bigg)-V_iV_j\Psi^{(4)}_{ij}-2V_i\Psi^{(5)}_{0i}\bigg]-\frac1{c^7}\bigg[\bigg(\frac32V^2+\varepsilon\bigg)\Psi^{(5)}_{00}
\\
&&\qquad+2V_i\Psi^{(6)}_{0i}+V_iV_j\Psi^{(5)}_{ij}\bigg]\bigg\}.
\een
and $\Sigma$ refers to the following expression
\ben\no
&&\Sigma=-\left[\rho V_i\frac{\partial U}{\partial x^i}-\frac{\partial pV_i}{\partial x^i}+U\left(\frac{\partial\rho}{\partial t}+\frac{\partial\rho V_i}{\partial x^i}\right)\right]+\frac{\rho}{c^2}\bigg[\bigg(2V^2+2U+\varepsilon+3\frac{p}\rho\bigg)\frac{\partial  U}{\partial t}-2V_i\frac{\partial\Phi}{\partial x^i}
\\\no
&&-V_i\frac{\partial \Pi_i}{\partial t}+\frac2\rho\frac{\partial pUV_i}{\partial x^i}\bigg]+\frac\rho{c^4}\bigg[\frac{\partial U}{\partial t}\bigg(2V^4+10UV^2-4U\varepsilon+2V^2\varepsilon+2V^2\frac{p}\rho-2\Pi_iV_i\bigg)+2\frac{\partial[pV_i(U^2+2\Phi)] }{\partial x^i}
\\\no
&&-V_j\frac{\partial U}{\partial x^j}\bigg(6U\varepsilon+6U\frac{p}\rho+8\Phi\bigg)+\bigg(\frac{\partial U}{\partial x^i}-\frac1\rho\frac{\partial p}{\partial x^i}\bigg)\left(\Psi^{(5)}_{0i}+3U^2V_i\right)+\frac{U}{4}\bigg(V^2+2\varepsilon+2\frac{p}\rho\bigg)\bigg(\frac{\partial \Psi^{(4)}_{kk}}{\partial t}+V_i\frac{\partial \Psi^{(4)}_{kk}}{\partial x^i}\bigg)
\\\no
&&-\left(2V^2+4\varepsilon\right)\frac{\partial \Phi}{\partial t}-V_i\frac{\partial \Phi}{\partial x^i}\bigg(4V^2+8U+6\varepsilon+6\frac{p}\rho\bigg)-V_i\frac{\partial \Pi_i}{\partial t}\bigg(V^2+4U+\varepsilon+\frac{p}\rho\bigg)+\frac12V_i\frac{\partial \Psi^{(6)}_{00}}{\partial x^i}
\\\no
&&+V_iV_j\bigg(\frac12\frac{\partial \Psi^{(5)}_{ij}}{\partial t}-\frac{\partial \Psi^{(5)}_{0j}}{\partial x^i}\bigg)\bigg]+\frac\rho{c^5}\bigg[\bigg(\frac{V^2}2+U+\varepsilon+\frac{p}{2\rho}\bigg)\frac{d\Psi^{(5)}_{00}}{dt}+V_iV_j\bigg(\frac{\partial\Psi^{(6)}_{0i}}{\partial x^j}-\frac12\frac{\partial\Psi^{(5)}_{ij}}{\partial t}\bigg)
\\\lb{bal3c}
&&-\frac{\partial pV_i}{\partial x^i}\Psi^{(5)}_{00}+\frac{\partial U}{\partial x^i}\left(V_i\Psi^{(5)}_{00}+\Psi^{(6)}_{0i}\right)+\frac{V_i}{2}\frac{\partial\Psi^{(7)}_{00}}{\partial x^i}\bigg].
\een

The total energy density of the gas $\rho e$ is the sum of its internal  and  kinetic   energy densities that in the Newtonian approximation are given by $\rho\varepsilon$ and $\rho V^2/2$, respectively. The    total energy density hydrodynamic equation is obtained from the  subtraction of the  continuity equation (\ref{bal1a}) multiplied by $c^2$ from the mass-energy hydrodynamic equation (\ref{bal2a}), which yields
\ben\lb{ba3d}
&&\frac{\partial\rho e}{\partial t}+\frac{\partial \rho e V_i}{\partial x^i}+\Sigma=0,
\een
where the total energy density is given by
\ben\no\lb{ba3c}
&&\rho e=\widetilde\sigma-\widetilde\rho=\rho\bigg\{\left(\frac{V^2}2+\varepsilon\right)+\frac1{c^2}\left[\frac58V^4+\frac52V^2U-\frac52U^2
+2U\varepsilon+V^2\left(\varepsilon+\frac{p}\rho\right)\right]+\frac1{c^4}\bigg[\frac{11}{16}V^6+\frac{47}8UV^4
\\\no
&&\qquad+\frac{23}4V^2U^2-\frac{U^3}2+\bigg(\varepsilon+\frac{p}\rho\bigg)\left(V^4+6UV^2\right)+4(V^2+\varepsilon)\Phi-\left(\frac32V^2+\varepsilon+\frac{p}\rho-U\right)\Pi_kV_k
\\
&&\qquad-\frac12V_iV_j\Psi^{(4)}_{ij}-V_i\Psi^{(5)}_{0i}\bigg]-\frac1{c^5}\left[\left(V^2+\varepsilon\right)\Psi^{(5)}_{00}
+V_i\Psi^{(6)}_{0i}+\frac12V_iV_j\Psi^{(5)}_{ij}\right]\bigg\}.
\een
Here we note that the hydrodynamic equation for the total energy density is known up to $1/c^5$ order terms.

The hydrodynamic equation for the momentum density refers to the space components of the conservation of energy-momentum tensor  and it follows by choosing  $\Upsilon=m^4 u^i$, namely
\ben\no
&&\frac{\partial \rho\widetilde{\frak{V}}_i}{\partial t}+\frac{\partial \rho\widetilde{\frak{V}}_iV_j}{\partial x^j}+\frac{\partial p}{\partial x^i}\left[1+\frac{2U}{c^2}-\frac1{c^4}\left(U^2+2\Phi+\frac{\Psi^{(4)}_{kk}}2\right)\right]
-\rho\frac{\partial U}{\partial x^i}\bigg\{1+\frac1{c^2}\bigg(2V^2+2U+\varepsilon+\frac{p}{\rho}\bigg)+\frac2{c^4}\bigg[V^4+5UV^2
\\\no
&&\qquad-\frac{3U^2}2+\Phi+(V^2+U)\bigg(\varepsilon+\frac{p}\rho\bigg)-\Pi_iV_i-\frac{\Psi^{(4)}_{kk}}4\bigg]\bigg\}
+\frac\rho{c^2}\bigg(V_j\frac{\partial \Pi_j}{\partial x^i}-2\frac{\partial \Phi}{\partial x^i}\bigg)\bigg[1+\frac1{c^2}\bigg(V^2+4U+\varepsilon+\frac{p}\rho\bigg)\bigg]
\\\lb{ba4h}
&&\qquad+\frac\rho{2c^4}\bigg[\frac{\partial \Psi^{(6)}_{00}}{\partial x^i}+2V_j\frac{\partial \Psi^{(5)}_{0j}}{\partial x^i}+V_jV_k\frac{\partial \Psi^{(4)}_{jk}}{\partial x^i}\bigg]+\frac\rho{c^5}\bigg[\frac{\Psi_{00}^{(5)}}2\left(\frac{\partial U}{\partial x^i}+\frac1\rho\frac{\partial p}{\partial x^i}\right)-\frac{\partial \Psi_{0i}^{(6)}}{\partial t}
+\frac12\frac{\partial \Psi_{00}^{(7)}}{\partial x^i}\bigg]=0,
\een
where the abbreviation for the momentum density was introduced
\ben\no
&&\rho\widetilde{\frak{V}}_i=\rho V_i\bigg\{1+\frac1{c^2}\bigg(V^2+6U+\varepsilon+\frac{p}\rho\bigg)+\frac1{c^4}\bigg[V^4+10V^2U
+13U^2+2\Phi+\big(V^2+6U\big)\bigg(\varepsilon+\frac{p}\rho\bigg)
\\\lb{ba4i}
&&\quad-2\Pi_iV_i-\frac{\Psi^{(4)}_{kk}}2\bigg]\bigg\}-\frac\rho{c^2}\Pi_i\bigg[1+\frac1{c^2}\bigg(V^2+4U+\varepsilon+\frac{p}\rho\bigg)\bigg]-\frac\rho{c^4}\left(\Psi^{(5)}_{0i}+\Psi^{(4)}_{ij}V_j\right)-\frac\rho{c^5}\left(\Psi^{(5)}_{ij}V_j+\Psi^{(5)}_{00}V_i\right).
\een
Note that like the hydrodynamic equation for the total energy density, the one for the momentum density is known up to $1/c^5$ order terms.

After some rearrangements, the above expressions (\ref{ba4h}) and  (\ref{ba4i}) correspond  to eqs. \emph{(54)} and \emph{(88)} of  the works \cite{Ch2,Ch4}, respectively.

\section{Conservation of total energy density}\lb{s.7}

According to Chandrasekhar \cite{Ch3}, to obtain the conserved energy in the $n$--th post-Newtonian approximation, we have to know the components of the metric tensor $g_{ij}$ in the $(n+1)$--th post-Newtonian approximation. In the present work,  we obtained the hydrodynamic equations  from a Boltzmann equation in the $2\frac12$--post-Newtonian approximation, and hence the conservation of total energy density refers to the one in the $1\frac12$--post-Newtonian approximation.

The energy density conservation law in the first post-Newtonian approximation was derived by Chandrasekhar\cite{Ch0} by requiring the isentropic flow condition, a consequence of the  first law of thermodynamics. In  subsequent works with some collaborators \cite{Ch1,Ch2,Ch3,Ch4,Ch5} the energy  density conservation law -- up to $2\frac12$-post-Newtonian approximation --  was derived by taking into account the component energy-momentum complex $\Theta^{00}$ subtracted from the mass density. For the latter, the terms $Q^{(3)}_{kk}/c^6$ and $Q^{(7)}_{kk}/c^7$ were included, which are contributions from the $3$ and $3 \frac12$ approximations, respectively.

Here we shall use the equation for the total energy density (\ref{ba3d}), which by considering terms up to $1/c^2$ and integrating over the volume occupied by the fluid reads 
\ben\no
\int_V\frac{\partial}{\partial t}\bigg\{\rho\left(\frac{V^2}2+\varepsilon\right)+\frac\rho{c^2}\left[\frac58V^4+\frac52V^2U-\frac52U^2
+2U\varepsilon+V^2\left(\varepsilon+\frac{p}\rho\right)\right]\bigg\}d^3x+\int_V\frac{\partial}{\partial x^i}\bigg\{\rho V_i\left(\frac{V^2}2+\varepsilon\right)
\\\no
+\frac{\rho V_i}{c^2}\left[\frac58V^4+\frac52V^2U-\frac52U^2
+2U\varepsilon+V^2\left(\varepsilon+\frac{p}\rho\right)\right]\bigg\}d^3x-\int_V\left[\rho V_i\frac{\partial U}{\partial x^i}-\frac{\partial pV_i}{\partial x^i}+U\left(\frac{\partial\rho}{\partial t}+\frac{\partial\rho V_i}{\partial x^i}\right)\right]d^3x
\\\lb{en1}
+\int_V\frac{\rho}{c^2}\bigg[\bigg(2V^2+2U+\varepsilon+3\frac{p}\rho\bigg)\frac{\partial  U}{\partial t}-2V_i\frac{\partial\Phi}{\partial x^i}
-V_i\frac{\partial \Pi_i}{\partial t}+\frac2\rho\frac{\partial pUV_i}{\partial x^i}\bigg]d^3x=0.
\een

The two first terms of (\ref{en1}) can be transformed into
\ben\lb{en2}
\frac{d}{dt}\int_V\bigg\{\rho\left(\frac{V^2}2+\varepsilon\right)+\frac\rho{c^2}\left[\frac58V^4+\frac52V^2U-\frac52U^2
+2U\varepsilon+V^2\left(\varepsilon+\frac{p}\rho\right)\right]\bigg\}d^3x,
\een
thanks to Reynolds'  transport theorem which is valid for an arbitrary scalar--, vector-- or tensor--valued function $F(\bx,t)$:
\ben
\frac{d}{dt}\int_VF(\bx,t)d^3x=\int_V\left\{\frac{\partial F(\bx,t)}{\partial t}+\frac{\partial F(\bx,t)V_i}{\partial x^i}\right\}d^3x.
\een

The sum of the two spatial divergence terms, namely 
\ben\lb{en3}
\int_V\left(\frac{\partial pV_i}{\partial x^i}+\frac2{c^2}\frac{\partial pUV_i}{\partial x^i}\right)d^3x=\int_Sp\left(1+\frac{2U}{c^2}\right)V_in_i dS=0,
\een
vanishes by using Gauss divergence theorem and by assuming that the pressure vanishes on the boundary of the configuration.

The following relationships
\ben\lb{en4a}
&&\int_V\left[\rho V_i\frac{\partial U}{\partial x^i}+U\left(\frac{\partial\rho}{\partial t}+\frac{\partial\rho V_i}{\partial x^i}\right)\right]d^3x=\frac12\frac{d}{dt}\int_V\rho Ud^3x,
\\\lb{en4b}
&&\int_V\frac{\rho}{c^2}\left[\bigg(2V^2+2U+\varepsilon+3\frac{p}\rho\bigg)\frac{\partial  U}{\partial t}-2V_i\frac{\partial\Phi}{\partial x^i}\right]d^3x=0,
\\\lb{en4c}
&&-\frac1{c^2}\int_V\rho V_i\frac{\partial\Pi_i}{\partial t}d^3x=-\frac1{2c^2}\frac{d}{dt}\int_V\rho V_i\Pi_id^3x,
\een
can be obtained after a long calculation by considering the definitions of the gravitational potentials (see \cite{Ch0,GK1})
\ben
U(\bx)=G\int_V\frac{\rho(\bx')}{\vert\bx-\bx'\vert}d^3x',\qquad U_i(\bx)=G\int_V\frac{\rho(\bx')V_i(\bx')}{\vert\bx-\bx'\vert}d^3x',\qquad
\Phi(\bx)=G\int_V\frac{\rho(\bx')\varphi(\bx')}{\vert\bx-\bx'\vert}d^3x', 
\\ \chi=-G\int_V\rho(\bx'){\vert\bx-\bx'\vert}d^3x',\qquad \varphi=V^2+U+\frac\varepsilon2+\frac{3p}{2\rho},\qquad
\Pi_i=4U_i-\frac12\frac{\partial^2\chi}{\partial t\partial x^i}.
\een

Now, by collecting the above expressions (\ref{en2}), (\ref{en3}), (\ref{en4a}) -- (\ref{en4c}), we get the conservation law  for the total energy density $\mathfrak{E}$ 
\ben
&&\frac{d}{dt}\int_V \mathfrak{E}\,d^3x=0,\qquad\hbox{where}
\\\lb{en5}
&&\mathfrak{E}=\rho\left(\frac{V^2}2+\varepsilon-\frac{U}2\right)+\frac\rho{c^2}\left[\frac58V^4+\frac52V^2U-\frac52U^2
+2U\varepsilon+V^2\left(\varepsilon+\frac{p}\rho\right)-\frac12\Pi_iV_i\right].
\een

\emph{Some remarks:} 
\begin{itemize}
\item[i)] The expression (\ref{en5}) represents the first-- and the $1\frac12$-- post-Newtonian approximations for the total energy density,  although no contribution of the gravitation potential $\Psi^{(5)}_{00}$, shows up. Indeed, the gravitational potential $\Psi^{(5)}_{00}$  appears only in the expressions  for the mass density $\widetilde\rho$ and for the mass-energy density $\widetilde\sigma$  in the $1/c^7$ terms. 
\item[ii)] In all approximations, the continuity equation can  always be derived, and according to Chandrasekhar  \cite{Ch0,Ch1,Ch2,Ch3,Ch4,Ch5} this reflects that in these higher approximations there is no alteration in the baryon number.
\item[iii)]  When  the terms $Q^{(3)}_{kk}/c^6$ and $Q^{(7)}_{kk}/c^7$ are included -- the contributions from the $3$ and $3 \frac12$ approximations --,  it was shown that the mean value of the energy rate in the $2\frac12$--post-Newtonian approximation "represents a secular decrease of the integrated energy of the system, a result that is in exact agreement with the rate of emission of gravitational radiant energy predicted by the linearized theory of gravitational radiation \cite{Ch4,Ch5}".
\end{itemize}

\section{Conclusions}\lb{s.8}

In this work we have analyzed a relativistic gas in the presence of gravitational fields within the framework of a kinetic theory described by a $2\frac12$--PN approximation. The Boltzmann equation in the $2\frac12$--PN approximation,  was obtained  from the evolution equation of the one-particle distribution function with respect to the proper time along the world line of a gas particle. The Boltzmann equation as well as the equilibrium Maxwell--J\"uttner distribution function were determined up to the $1/c^7$--order as functions of the Newtonian gravitation potential $U$,  of the scalar gravitational potentials $\Phi$,  $\Psi^{(5)}_{00},$ $\Psi^{(6)}_{00}$, $\Psi^{(7)}_{00}$,  of the vector gravitational potentials $\Pi_i$, $\Psi^{(5)}_{0i}$, $\Psi^{(6)}_{0i}$, and of the tensor gravitational potentials  $\Psi^{(4)}_{ij}$, $\Psi^{(5)}_{ij}$. The expression for the Boltzmann equation was derived from  knowledge of the components of the Christoffel symbols, which were checked with the values given in the literature. The Maxwell--J\"uttner distribution function  was determined from symbolic computation. The components of the particle four-flow and energy-momentum tensor were calculated by symbolic computation from the equilibrium Maxwell--J\"uttner distribution function and correspond to those of the phenomenological equations, which makes use  of their decompositions in terms of the hydrodynamic four-velocity. The hydrodynamic equations for the mass, mass-energy and momentum densities -- which represent the Eulerian equations for a relativistic gas in the presence of gravitational fields -- were obtained by symbolic computation from the Boltzmann equation, which matches the corresponding equations derived from the conservation equations of the particle four-flow and energy-momentum tensor. We observe that the production term of the Boltzmann equation vanishes for the quantities related with the rest mass and momentum four-vector of the gas particles.   The hydrodynamic equation for the total energy density is derived from the corresponding equations for the  mass and the mass-energy, which leads to the total energy conservation law in the $1\frac12$--PN approximation, since we have to know the components of the metric tensor $g_{ij}$ in the $3\frac12$--th post-Newtonian approximation to obtain the energy conservation law in the $2\frac12$--th post-Newtonian approximation. As a final remark we comment about the orders of the equations obtained from the Boltzmann equation and Maxwell--J\"uttner distribution function. The results that follow show that the components of the energy-momentum tensor and of the balance equations for the mass-energy and momentum densities are known up to the $1/c^5$--order. However, the expression for the mass density is known up to the $1/c^7$--order, which by  multiplication with $c^2$ results  in an equation of the $1/c^5$--order  where the first term represents the energy density associated with the mass density $\rho c^2.$
\vfill

\newpage
 
{\bf Appendix A}

The Christoffel symbols corresponding to the  metric tensor components in the $2\frac12$-- post-Newtonian approximation are 

\begin{tiny}
\ben\no
&&{{\Gamma^0}_{00}}=-\frac1{c^3}\frac{\partial U}{\partial t}+\frac1{c^5}\left(\Pi_i\frac{\partial U}{\partial x^i}-2\frac{\partial \Phi}{\partial t}\right)+\frac1{2c^6}\frac{d\Psi^{(5)}_{00}}{dt}+\frac1{c^7}\bigg[\frac12\frac{\partial\Psi^{(6)}_{00}}{\partial t}+2U^2\frac{\partial U}{\partial t}-4\left(U\frac{\partial \Phi}{\partial t}+\Phi\frac{\partial U}{\partial t}\right)+{\Pi_j}\frac{\partial \Pi_j}{\partial t}
\\\no
&&\qquad-\Pi_j\frac{\partial (U^2-2\Phi)}{\partial x^j}+\Psi^{(5)}_{0j}\frac{\partial U}{\partial x^j}\bigg]+\frac1{c^8}\left[\frac12\frac{\partial\Psi^{(7)}_{00}}{\partial t}+U\frac{\partial \Psi^{(5)}_{00}}{\partial t}+\Psi^{(5)}_{00}\frac{\partial U}{\partial t}+\Psi^{(6)}_{0i}\frac{\partial U}{\partial x^i}\right],
\\\no
&&{{\Gamma^0}_{0i}}=-\frac1{c^2}\frac{\partial U}{\partial x^i}-\frac2{c^4}\frac{\partial\Phi}{\partial x^i}+\frac1{c^6}\left[\frac12\frac{\partial \Psi^{(6)}_{00}}{\partial x^i}+2U^2\frac{\partial U}{\partial x^i}-4\left(U\frac{\partial \Phi}{\partial x^i}+\Phi\frac{\partial U}{\partial x^i}\right)+\frac{\Pi_j}{2}\left(\frac{\partial \Pi_j}{\partial x^i}-\frac{\partial \Pi_i}{\partial x^j}-2\frac{\partial U}{\partial t}\delta_{ij}\right)\right]
\\\no
&&\qquad+\frac1{c^7}\left[\frac12\frac{\partial \Psi^{(7)}_{00}}{\partial x^i}+\Psi^{(5)}_{00}\frac{\partial U}{\partial x^i}\right],
\\\no
&&{{\Gamma^i}_{00}}=-\frac1{c^2}\frac{\partial U}{\partial x^i}+\frac1{c^4}\left[2\frac{\partial (U^2-\Phi)}{\partial x^i}-\frac{\partial\Pi_i}{\partial t}\right]+\frac1{c^6}\left[\frac12\frac{\partial\Psi^{(6)}_{00}}{\partial x^i}-\frac{\partial\Psi^{(5)}_{0i}}{\partial t}-\Pi_i\frac{\partial U}{\partial t}-4U\bigg(\frac{\partial U^2}{\partial x^i}-\frac{\partial \Phi}{\partial x^i}\bigg)
+2U\frac{\partial \Pi_i}{\partial t}-\Psi^{(4)}_{ij}\frac{\partial U}{\partial x^j}\right]
\\\no
&&\qquad+\frac1{2c^7}\left[\frac{\partial\Psi^{(7)}_{00}}{\partial x^i}-2\frac{\partial\Psi^{(6)}_{0i}}{\partial t}-2\Psi^{(5)}_{ij}\frac{\partial U}{\partial x^j}\right]+\frac1{c^8}\left[\Pi_i\frac{\partial(U^2-2\Phi)}{\partial t}-\Psi^{(5)}_{0i}\frac{\partial U}{\partial t}+U\left(2\frac{\partial\Psi^{(5)}_{0i}}{\partial t}-\frac{\partial\Psi^{(6)}_{00}}{\partial x^i}\right)\right.
\\\no
&&\left.\qquad-\left(\Psi^{(4)}_{ij}+4U^2\delta_{ij}\right)\left(\frac{\partial \Pi_j}{\partial t}-\frac{\partial(U^2-2\Phi)}{\partial x^j}\right)\right]+\frac1{2c^9}\left[\Pi_i\frac{\partial\Psi^{(5)}_{00}}{\partial t}+2U\left(2\frac{\partial\Psi^{(6)}_{0i}}{\partial t}-\frac{\partial\Psi^{(7)}_{00}}{\partial x^i}\right)\right.
\\\no
&&\left.\qquad-2\Psi^{(5)}_{ij}\left(\frac{\partial \Pi_j}{\partial t}-\frac{\partial(U^2-2\Phi)}{\partial x^j}\right)\right],
\\\no
&&{{\Gamma^0}_{ij}}=\frac1{2c^3}\left(\frac{\partial\Pi_i}{\partial x^j}+\frac{\partial\Pi_j}{\partial x^i}+2\frac{\partial U}{\partial t}\delta_{ij}\right)+\frac1{c^5}\left[U\left(\frac{\partial \Pi_i}{\partial x^j}+\frac{\partial \Pi_j}{\partial x^i}+2\delta_{ij}\frac{\partial U}{\partial t}\right)-\left(\Pi_i\frac{\partial U}{\partial x^j}+\Pi_j\frac{\partial U}{\partial x^i}-\Pi_k\frac{\partial U}{\partial x^k}\delta_{ij}\right)\right.
\\\no
&&\left.\qquad+\frac12\left(\frac{\partial \Psi^{(5)}_{0i}}{\partial x^j}+\frac{\partial \Psi^{(5)}_{0j}}{\partial x^i}-\frac{\partial \Psi^{(4)}_{ij}}{\partial t}\right)\right]+\frac1{2c^6}\left[\frac{\partial \Psi^{(6)}_{0i}}{\partial x^j}+\frac{\partial \Psi^{(6)}_{0j}}{\partial x^i}-\frac{\partial \Psi^{(5)}_{ij}}{\partial t}\right],
\\\no
&&{{\Gamma^i}_{0j}}=\frac1{2c^3}\left(\frac{\partial\Pi_j}{\partial x^i}-\frac{\partial\Pi_i}{\partial x^j}+2\frac{\partial U}{\partial t}\delta_{ij}\right)-\frac1{c^5}\left[\Pi_i\frac{\partial U}{\partial x^j}-U\left(\frac{\partial\Pi_i}{\partial x^j}-\frac{\partial \Pi_j}{\partial x^i}-2\frac{\partial U}{\partial t}\delta_{ij}\right)+\frac1{2}\left(\frac{\partial \Psi^{(5)}_{0i}}{\partial x^j}-\frac{\partial \Psi^{(5)}_{0j}}{\partial x^i}+\frac{\partial\Psi^{(4)}_{ij}}{\partial t}\right)\right]
\\\no
&&\qquad-\frac1{2c^6}\frac{d\Psi^{(5)}_{ij}}{dt}+\frac1{c^7}\left[\Pi_i\frac{\partial (U^2-2\Phi)}{\partial x^j}-\frac12\left(\Psi^{(4)}_{ik}+4U^2\delta_{ik}\right)\left(\frac{\partial\Pi_k}{\partial x^j}-\frac{\partial \Pi_j}{\partial x^k}-2\frac{\partial U}{\partial t}\delta_{jk}\right)+U\left(\frac{\partial \Psi^{(5)}_{0i}}{\partial x^j}-\frac{\partial \Psi^{(5)}_{0j}}{\partial x^i}+\frac{\partial\Psi^{(4)}_{ij}}{\partial t}\right)\right.
\\\no
&&\qquad\left.-\Psi^{(5)}_{0i}\frac{\partial U}{\partial x^j}\right]+\frac1{c^8}\left[U\frac{\partial \Psi^{(5)}_{ij}}{\partial t}-\Psi^{(6)}_{0i}\frac{\partial U}{\partial x^j}-\frac{\Psi^{(5)}_{ik}}2\left(\frac{\partial\Pi_k}{\partial x^j}-\frac{\partial \Pi_j}{\partial x^k}-2\frac{\partial U}{\partial t}\delta_{jk}\right)\right],
\\\no
&&{{\Gamma^i}_{jk}}=\frac1{c^2}\left(\frac{\partial U}{\partial x^j}\delta_{ik}+\frac{\partial U}{\partial x^k}\delta_{ij}-\frac{\partial U}{\partial x^i}\delta_{jk}\right)-\frac1{c^4}\left[\frac12\left(\frac{\partial\Psi^{(4)}_{ij}}{\partial x^k}+\frac{\partial\Psi^{(4)}_{ik}}{\partial x^j}-\frac{\partial\Psi^{(4)}_{jk}}{\partial x^i}\right)
+\left(\frac{\partial U^2}{\partial x^k}\delta_{ij}+\frac{\partial U^2}{\partial x^j}\delta_{ik}-\frac{\partial U^2}{\partial x^i}\delta_{jk}\right)\right]
\\\no
&&\qquad+\frac1{c^6}\left[\frac{\Pi_i}{2}\left(\frac{\partial\Pi_k}{\partial x^j}+\frac{\partial\Pi_j}{\partial x^k}+2\frac{\partial U}{\partial t}\delta_{jk}\right)+U\left(\frac{\partial\Psi^{(4)}_{ij}}{\partial x^k}+\frac{\partial\Psi^{(4)}_{ik}}{\partial x^j}-\frac{\partial\Psi^{(4)}_{jk}}{\partial x^i}\right)+\left(\Psi^{(4)}_{il}+4U\delta_{il}\right)\left(\frac{\partial U}{\partial x^k}\delta_{jl}+\frac{\partial U}{\partial x^j}\delta_{kl}-\frac{\partial U}{\partial x^l}\delta_{jk}\right)\right]
\\\no
&&\qquad+\frac1{c^7}\left[\frac{\partial U\Psi^{(5)}_{ij}}{\partial x^k}+\frac{\partial U\Psi^{(5)}_{ik}}{\partial x^j}-\frac{\partial U\Psi^{(5)}_{jk}}{\partial x^i}+\frac{\partial U}{\partial x^l}\left(\Psi^{(5)}_{jk}\delta_{li}-\Psi^{(5)}_{il}\delta_{jk}\right)\right].
\een
\end{tiny}

\vfill
\pagebreak

{\bf Appendix B}

The expressions for the Maxwell--J\"uttner distribution function in the $1/c^6$ and $1/c^7$ orders are:
\begin{tiny}
\ben\no
&&f^{(6)}=\Bigg\{-\frac{9 m^3 W^{12}}{1024 {(kT)}^3}+\frac{15 m^2 W^{10}}{128 (kT)^2}-\frac{9 m^3 U W^{10}}{64 (kT)^3}-\frac{9 m^3 (VW) W^{10}}{128 (kT)^3}-\frac{9 m^3 V^2 W^{10}}{256 ({kT})^3}+\frac{7 m^2 V^2 W^8}{16 ({kT})^2}+\frac{7 m^2 U W^8}{4 ({kT})^2}
\\\no
&&+\frac{7 m^2 ({VW}) W^8}{8 (kT)^2}-\frac{415 m W^8}{1024 (kT)}-\frac{3 m^3 U^2 W^8}{4 (kT)^3}-\frac{3 m^3 U (VW) W^8}{4 (kT)^3}-\frac{3 m^3 U V^2 W^8}{8 (kT)^3}-\frac{3 m^3 V^2 (VW) W^8}{16 (kT)^3}-\frac{3 m^3 V^4 W^8}{64 (kT)^3}
\\\no
&&-\frac{57 m^3 (VW)^2 W^8}{256 (kT)^3}+\frac{9 m^2 V^4 W^6}{16 (kT)^2}+\frac{57 m^2 U^2 W^6}{8 (kT)^2}+\frac{9 m^2 U V^2 W^6}{2 (kT)^2}+\frac{5 m^2 (VW)^2 W^6}{2 (kT)^2}+\frac{9 m^2 V^2 (VW) W^6}{4 (kT)^2}+\frac{9 m^2 U (VW) W^6}{(kT)^2}
\\\no
&&+\frac{3 m^2 \Phi  W^6}{4 (kT)^2}+\frac{75 W^6}{128}-\frac{165 m U W^6}{32 (kT)}-\frac{165 m (VW) W^6}{64 (kT)}-\frac{165 m V^2 W^6}{128 (kT)}-\frac{3 m^2(\Pi V) W^6}{8 (kT)^2}-\frac{3 m^2(\Pi W) W^6}{8 (kT)^2}-\frac{m^3 U^2 V^2 W^6}{(kT)^3}
\\\no
&&-\frac{2 m^3 U^2 (VW) W^6}{(kT)^3}-\frac{m^3 U V^2 (VW) W^6}{(kT)^3}-\frac{4 m^3 U^3 W^6}{3 (kT)^3}-\frac{m^3 U V^4 W^6}{4 (kT)^3}-\frac{11 m^3 U (VW)^2 W^6}{8 (kT)^3}-\frac{m^3 V^4 (VW) W^6}{8 (kT)^3}
\\\no
&&-\frac{11 m^3 V^2 (VW)^2 W^6}{32 (kT)^3}-\frac{m^3 V^6 W^6}{48 (kT)^3}-\frac{17 m^3 (VW)^3 W^6}{48 (kT)^3}+\frac{m^2 V^6 W^4}{4 (kT)^2}+\frac{3 m^2 U V^4 W^4}{(kT)^2}+\frac{6 m^2 U^3 W^4}{(kT)^2}+\frac{27 m^2 (VW)^3 W^4}{8 (kT)^2}
\\\no
&&+\frac{19 m^2 U^2 V^2 W^4}{2 (kT)^2}+\frac{45 V^2 W^4}{32}+\frac{31 m^2 V^2 (VW)^2 W^4}{8 (kT)^2}+\frac{31 m^2 U (VW)^2 W^4}{2 (kT)^2}+\frac{45 U W^4}{8}+\frac{3 m^2 V^4 (VW) W^4}{2 (kT)^2}+\frac{19 m^2 U^2 (VW) W^4}{(kT)^2}
\\\no
&&+\frac{12 m^2 U V^2 (VW) W^4}{(kT)^2}+\frac{45 (VW) W^4}{16}+\frac{3 m(\Pi V) W^4}{2 (kT)}+\frac{3 m(\Pi W) W^4}{2 (kT)}+\frac{m^2 V^2 \Phi  W^4}{(kT)^2}+\frac{4 m^2 U \Phi  W^4}{(kT)^2}+\frac{2 m^2 (VW) \Phi  W^4}{(kT)^2}
\\\no
&&-\frac{3 m \Phi  W^4}{(kT)}-\frac{57 m U^2 W^4}{4 (kT)}-\frac{87 m U (VW) W^4}{4 (kT)}-\frac{87 m U V^2 W^4}{8 (kT)}-\frac{87 m V^2 (VW) W^4}{16 (kT)}-\frac{87 m V^4 W^4}{64 (kT)}-\frac{765 m (VW)^2 W^4}{128 (kT)}
\\\no
&&-\frac{2 m^2 U(\Pi V) W^4}{(kT)^2}-\frac{m^2 (VW)(\Pi V) W^4}{(kT)^2}-\frac{2 m^2 U(\Pi W) W^4}{(kT)^2}-\frac{m^2 V^2(\Pi V) W^4}{2 (kT)^2}-\frac{m^2 V^2(\Pi W) W^4}{2 (kT)^2}-\frac{11 m^2 (VW)(\Pi W) W^4}{8 (kT)^2}
\\\no
&&-\frac{3 m^2W_iW_j\Psi^{(4)}_{ij} W^4}{16 (kT)^2}-\frac{m^3 U (VW)^3 W^4}{(kT)^3}-\frac{m^3 U^2 (VW)^2 W^4}{(kT)^3}-\frac{m^3 U V^2 (VW)^2 W^4}{2 (kT)^3}-\frac{m^3 V^2 (VW)^3 W^4}{4 (kT)^3}-\frac{m^3 V^4 (VW)^2 W^4}{16 (kT)^3}
\\\no
&&-\frac{19 m^3 (VW)^4 W^4}{64 (kT)^3}-\frac{1035 (kT) W^4}{1024 m}+\frac{15 V^4 W^2}{16}+\frac{17 m^2 (VW)^4 W^2}{8 (kT)^2}+\frac{5 m^2 V^2 (VW)^3 W^2}{2 (kT)^2}+\frac{10 m^2 U (VW)^3 W^2}{(kT)^2}+\frac{45 U^2 W^2}{8}
\\\no
&&+\frac{15}{2} U V^2 W^2+\frac{3 m^2 V^4 (VW)^2 W^2}{4 (kT)^2}+\frac{19 m^2 U^2 (VW)^2 W^2}{2 (kT)^2}+\frac{6 m^2 U V^2 (VW)^2 W^2}{(kT)^2}+\frac{135 (VW)^2 W^2}{32}+\frac{15}{4} V^2 (VW) W^2
\\\no
&&+15 U (VW) W^2+\frac{2 m V^2(\Pi V) W^2}{(kT)}+\frac{6 m U(\Pi V) W^2}{(kT)}+\frac{4 m (VW)(\Pi V) W^2}{(kT)}+\frac{2 m V^2(\Pi W) W^2}{(kT)}+\frac{6 m U(\Pi W) W^2}{(kT)}+\frac{9 m (VW)(\Pi W) W^2}{2 (kT)}
\\\no
&&+\frac{m^2 (VW)^2 \Phi  W^2}{(kT)^2}+\frac{15 \Phi  W^2}{4}+\frac{mV_iV_j\Psi^{(4)}_{ij} W^2}{2 (kT)}+\frac{mV_jW_i\Psi^{(4)}_{ij} W^2}{2 (kT)}+\frac{mV_iW_j\Psi^{(4)}_{ij} W^2}{2 (kT)}+\frac{3 mW_iW_j\Psi^{(4)}_{ij} W^2}{4 (kT)}
\\\no
&&+\frac{m V_i \text{$\Psi^{(5)}_{0i} $} W^2}{{kT}}+\frac{mW_i\Psi^{(5)}_{0i} W^2}{(kT)}+\frac{m\Psi^{(6)}_{00} W^2}{2 (kT)}-\frac{15(\Pi V) W^2}{8}-\frac{15(\Pi W) W^2}{8}-\frac{6 m U V^4 W^2}{(kT)}-\frac{2 m U^3 W^2}{(kT)}
\\\no
&&-\frac{14 m U^2 V^2 W^2}{(kT)}-\frac{3 m V^4 (VW) W^2}{(kT)}-\frac{28 m U^2 (VW) W^2}{(kT)}-\frac{24 m U V^2 (VW) W^2}{(kT)}-\frac{4 m V^2 \Phi  W^2}{(kT)}-\frac{12 m U \Phi  W^2}{(kT)}
\\\no
&&-\frac{8 m (VW) \Phi  W^2}{(kT)}-\frac{m V^6 W^2}{2 (kT)}-\frac{231 m U (VW)^2 W^2}{8 (kT)}-\frac{95 m (VW)^3 W^2}{16 (kT)}-\frac{231 m V^2 (VW)^2 W^2}{32 (kT)}-\frac{2 m^2 U (VW)(\Pi W) W^2}{(kT)^2}
\\\no
&&-\frac{m^2 UW_iW_j\Psi^{(4)}_{ij} W^2}{(kT)^2}
-\frac{m^2 (VW)^2(\Pi V) W^2}{2 (kT)^2}-\frac{3 m^2 (VW)^2(\Pi W) W^2}{2 (kT)^2}-\frac{m^2 V^2 (VW)(\Pi W) W^2}{2 (kT)^2}-\frac{m^2 (VW)W_iW_j\Psi^{(4)}_{ij} W^2}{2 (kT)^2}
\\\no
&&-\frac{m^2 V^2W_iW_j\Psi^{(4)}_{ij} W^2}{4 (kT)^2}-\frac{m^3 U (VW)^4 W^2}{4 (kT)^3}-\frac{m^3 (VW)^5 W^2}{8 (kT)^3}-\frac{m^3 V^2 (VW)^4 W^2}{16 (kT)^3}-\frac{345 (kT) U W^2}{64 m}
\\\no
&&-\frac{345 (kT) (VW) W^2}{128 m}-\frac{345 (kT) V^2 W^2}{256 m}+\frac{m^2 (VW)^5}{2 (kT)^2}+\frac{m^2 V^2 (VW)^4}{2 (kT)^2}+\frac{2 m^2 U (VW)^4}{(kT)^2}+\frac{15 (VW)^3}{8}+\frac{15 V^2 (VW)^2}{8}
\\\no
&&+\frac{15 U (VW)^2}{2}+\frac{2 m (VW)^2(\Pi V)}{(kT)}+\frac{3 m (VW)^2(\Pi W)}{(kT)}+\frac{2 m V^2 (VW)(\Pi W)}{(kT)}+\frac{6 m U (VW)(\Pi W)}{(kT)}+\frac{m V^2V_iV_j\Psi^{(4)}_{ij}}{(kT)}
\\\no
&&+\frac{mV_iV_j (VW)\Psi^{(4)}_{ij}}{(kT)}+\frac{m V^2V_jW_i\Psi^{(4)}_{ij}}{2 (kT)}+\frac{mV_j (VW)W_i\Psi^{(4)}_{ij}}{(kT)}+\frac{m V^2V_iW_j\Psi^{(4)}_{ij}}{2 (kT)}+\frac{mV_i (VW)W_j\Psi^{(4)}_{ij}}{(kT)}+\frac{m UW_iW_j\Psi^{(4)}_{ij}}{(kT)}
\\\no
&&+\frac{3 m (VW)W_iW_j\Psi^{(4)}_{ij}}{2 (kT)}+\frac{2 m V^2V_i\Psi^{(5)}_{0i}}{(kT)}+\frac{2 mV_i (VW)\Psi^{(5)}_{0i}}{(kT)}+\frac{3 m (VW)W_i\Psi^{(5)}_{0i}}{2 (kT)}-\frac{15 (VW)(\Pi W)}{8}-\frac{15W_iW_j\Psi^{(4)}_{ij}}{16}
\\\no
&&-\frac{3 m V^2 (VW)^3}{(kT)}-\frac{12 m U (VW)^3}{(kT)}-\frac{14 m U^2 (VW)^2}{(kT)}-\frac{12 m U V^2 (VW)^2}{(kT)}-\frac{4 m (VW)^2 \Phi }{(kT)}-\frac{mV_iV_j \text{V}^2\Psi^{(4)}_{ij}}{(kT)}-\frac{mV_iV_j (VW) \Psi^{(4)}_{ij}}{(kT)}
\\\no
&&-\frac{mV_i \text{V}^2W_j\Psi^{(4)}_{ij}}{(kT)}-\frac{m V^2W_i\Psi^{(5)}_{0i}}{(kT)}-\frac{2 mV_i \text{V}^2 \Psi^{(5)}_{0i}}{(kT)}-\frac{mV_i (VW) \Psi^{(5)}_{0i}}{(kT)}-\frac{3 m V^4 (VW)^2}{2 (kT)}-\frac{m(\Pi W)^2}{2 (kT)}-\frac{m (VW)W_i\Psi^{(5)}_{0i}}{2 (kT)}
\\\no
&&-\frac{mV_jW_i (VW)\Psi^{(4)}_{ij}}{4 (kT)}-\frac{mV_iW_j (VW)\Psi^{(4)}_{ij}}{4 (kT)}-\frac{mV_i \text{W}^2W_j \Psi^{(4)}_{ij}}{4 (kT)}-\frac{mV_jW_i \text{W}^2 \Psi^{(4)}_{ij}}{4 (kT)}-\frac{135 m (VW)^4}{64 (kT)}
\\\no
&&-\frac{m^2 (VW)^3(\Pi W)}{2 (kT)^2}-\frac{m^2 (VW)^2W_iW_j\Psi^{(4)}_{ij}}{4 (kT)^2}-\frac{m^3 (VW)^6}{48 (kT)^3}-\frac{345 (kT) (VW)^2}{256 m}-\frac{3285 (kT)^3}{1024 m^3}\Bigg\},
\een
\end{tiny}

\vfill

\begin{tiny}
\ben\no
&&f^{(7)}=\Bigg\{-\frac{m^2 U W^4 \Psi^{(5)}_{00}}{(kT)^2}-\frac{m^2 U W^2 W_iW_j\Psi^{(5)}_{ij}}{(kT)^2}-\frac{m^2 V^2 W^4 \Psi^{(5)}_{00}}{4 (kT)^2}-\frac{m^2 V^2 W^2 W_iW_j\Psi^{(5)}_{ij}}{4 (kT)^2}-\frac{m^2 VW^2 W^2 \Psi^{(5)}_{00}}{4 (kT)^2}
\\\no
&&-\frac{m^2 VW^2 W_iW_j\Psi^{(5)}_{ij}}{4 (kT)^2}-\frac{m^2 VW W^4 \Psi^{(5)}_{00}}{2 (kT)^2}-\frac{m^2 VW W^2 W_iW_j\Psi^{(5)}_{ij}}{2 (kT)^2}-\frac{3 m^2 W^6 \Psi^{(5)}_{00}}{16 (kT)^2}-\frac{3 m^2 W^4 W_iW_j\Psi^{(5)}_{ij}}{16 (kT)^2}
\\\no
&&+\frac{3 m U W^2 \Psi^{(5)}_{00}}{(kT)}+\frac{m U W_iW_j\Psi^{(5)}_{ij}}{(kT)}+\frac{2 m V^2 V_i\Psi^{(6)}_{0i}}{(kT)}-\frac{2 m V^2 V_i\Psi^{(6)}_{0i}}{(kT)}+\frac{m V^2 W^2 \Psi^{(5)}_{00}}{(kT)}+\frac{m V^2 W_iW_j\Psi^{(5)}_{ij}}{2 (kT)}
\\\no
&&-\frac{m V^2 W_i\Psi^{(6)}_{0i}}{(kT)}+\frac{m V_iV_jW^2 \Psi^{(5)}_{ij}}{2 (kT)}+\frac{m V_iVW W_j\Psi^{(5)}_{ij}}{2 (kT)}+\frac{2 m V_iVW \Psi^{(6)}_{0i}}{(kT)}-\frac{m V_iVW \Psi^{(6)}_{0i}}{(kT)}+\frac{m V_iW^2 W_j\Psi^{(5)}_{ij}}{2 (kT)}
\\\no
&&+\frac{m V_iW^2 \Psi^{(6)}_{0i}}{(kT)}+\frac{m V_jVW W_i\Psi^{(5)}_{ij}}{2 (kT)}+\frac{m V_jW^2 W_i\Psi^{(5)}_{ij}}{2 (kT)}+\frac{m VW^2 \Psi^{(5)}_{00}}{(kT)}+\frac{2 m VW W^2 \Psi^{(5)}_{00}}{(kT)}+\frac{m VW W_iW_j\Psi^{(5)}_{ij}}{(kT)}
\\\no
&&+\frac{3 m VW W_i\Psi^{(6)}_{0i}}{2 (kT)}-\frac{m VW W_i\Psi^{(6)}_{0i}}{2 (kT)}+\frac{3 m W^4 \Psi^{(5)}_{00}}{4 (kT)}+\frac{3 m W^2 W_iW_j\Psi^{(5)}_{ij}}{4 (kT)}+\frac{m W^2 W_i\Psi^{(6)}_{0i}}{(kT)}+\frac{m W^2 \Psi^{(7)}_{00}}{2 (kT)}
\\\no
&&-\frac{15 W^2 \Psi^{(5)}_{00}}{16}-\frac{15 W_iW_j\Psi^{(5)}_{ij}}{16}\Bigg\}.
\een
\end{tiny}

\end{document}